\documentclass[11pt,aps,prd,nofootinbib,superscriptaddress,preprint,preprintnumbers
]{revtex4-2}

\pdfoutput=1 

\usepackage{amssymb}
\usepackage{amsmath}
\usepackage[utf8]{inputenc}
\usepackage{autobreak}
\usepackage{array,hhline}
\usepackage{bbm}
\usepackage{bm}
\usepackage{booktabs}
\usepackage{bbding}
\usepackage{caption}
\usepackage{cancel}
\usepackage{float}
\usepackage{graphicx}
\usepackage{hyperref}
\usepackage{multirow}
\usepackage{makecell}
\usepackage{pgffor}
\usepackage{subfig}
\usepackage{textcomp}
\usepackage{upgreek,textgreek}
\usepackage{xcolor}

\newcommand{\SKLP}{State Key Laboratory of Particle Detection and Electronics, University of Science and Technology of China, Hefei 230026, Anhui, People’s Republic of China}
\newcommand{\USTC}{Department of Modern Physics, University of Science and Technology of China, Hefei 230026, Anhui, People’s Republic of China}
\newcommand{\NUS}{Department of Physics, National University of Singapore, Singapore 117551, Singapore}

\begin{document}

\title{\boldmath Mixed $\text{QCD} \otimes \text{EW}$ corrections to charged Higgs pair production in THDM at electron-positron colliders}

\author{Zhi-Xing Zhang}
\affiliation{\SKLP}
\affiliation{\USTC}

\author{Ren-You Zhang}
\email{zhangry@ustc.edu.cn}
\affiliation{\SKLP}
\affiliation{\USTC}

\author{Zhe Li}
\affiliation{\SKLP}
\affiliation{\USTC}

\author{Shu-Xiang Li}
\affiliation{\SKLP}
\affiliation{\USTC}

\author{Wen-Jie He}
\affiliation{\SKLP}
\affiliation{\USTC}

\author{Liang Han}
\affiliation{\SKLP}
\affiliation{\USTC}

\author{Qing-hai Wang}
\affiliation{\NUS}

\begin{abstract}
We calculate the two-loop mixed QCD$\otimes$EW corrections for the charged Higgs boson pair production within the framework of four types of Two Higgs Doublet Models (THDMs) with the $Z_2$ symmetry. We analyze in detail the dependences of our results on physical parameters, including the charged Higgs mass, $\tan\beta$, the scattering angle, and the colliding energy. It is noticeable that the mixed QCD$\otimes$EW relative correction is independent of the scattering angle due to the topology of Feynman diagrams at $O(\alpha\alpha_s)$. Numerical results in most allowed regions of four types of THDMs are provided in the density plots on the $m_{H^{\pm}}$-$\tan\beta$ plane.
For type-I and type-X, the mixed QCD$\otimes$EW relative correction varies slightly near $1\%$ except in the vicinity of resonance. For type-II and type-Y, the corrections increase consistently in large $\tan\beta$ region and reach up to $11.5\%$ at $\tan\beta = 50$. We also compute the $O(\alpha)$ corrections to obtain the corrected cross section up to $O(\alpha\alpha_s)$. The numerical results show that the corrected cross section can be larger than $80\ \mathrm{fb}$ in some parameter space region for type-I and type-X THDMs.

\begin{description}
\item[keywords]
NNLO QCD$\otimes$EW corrections, THDM, charged Higgs boson
\end{description}
\end{abstract}

\maketitle

\section{Introduction}
\label{sec:intro}
\par
The Standard Model (SM) particle spectrum was completed on July 4th, 2012, with the 125 GeV Higgs boson discovered by the ATLAS \cite{ATLAS:2012yve} and CMS \cite{CMS:2012qbp} collaborations at the Large Hadron Collider (LHC). Therefore, the SM is thus far regarded as the most successful framework of fundamental particles. Despite no signal for new physics beyond the Standard Model (BSM) discovered from current experimental datasets, the SM is unable to answer several phenomenological conundrums of the observable universe, including the nature of dark matter, the origin of neutrino masses, the problem of baryon asymmetry, and so forth. In addition, the SM involves theoretical drawbacks such as the hierarchy problem, no unified description for the gauge group and the flavor structure. These problems hint at new physics, and we can regard the SM as an effective approximation of a more fundamental framework in the low energy scenario. Thus, extending the SM is a reasonable way to solve these problems and find new physics. Among multiple SM extensions, the Two Higgs Doublet Model (THDM) is appealing as one of the simplest extensions of the Higgs sector \cite{Gunion:1989we,Lee:1973iz,Gunion:2002zf,Branco:2011iw}. The THDM can also be regarded as the scalar sector of many other complex and fundamental models, such as supersymmetric models (SUSY) \cite{Haber:1984rc} and axion models \cite{Kim:1986ax,Peccei:1977hh}. Furthermore, the rich phenomenology the THDM provides, including charged Higgs bosons, spontaneous $\mathcal{CP}$-violation \cite{Turok:1990zg,Davies:1994id,Cline:1995dg,Fromme:2006cm} and dark matter candidates \cite{Deshpande:1977rw,Dolle:2009fn}, helps to search BSM signals. Consequently, the THDM can serve as a tool for studying the Higgs sector of new physics models, free from parameter constraints in various fundamental theories, making it more convenient and straightforward.

\par
Searching for new fundamental particles is one of the most direct approaches to exploring BSM physics. The Higgs sector predicted by the THDM comprises five physical Higgs bosons, including $h^0$, $H^0$, $A^0$ and $H^{\pm}$. Since the SM Higgs boson is neutral, the discovery of a charged Higgs boson would conclusively signify the existence of BSM new physics, hence searches for such particles have been a priority in experiments and great efforts have been taken. In the low mass region, $t\rightarrow H^+b$ and $H^+\rightarrow \tau^+\nu$ and $H^+\rightarrow \tau^+\nu$ are respectively the dominant production channel and decay mode of the charged Higgs boson. These two processes have been studied at the LEP \cite{ALEPH:2000gfn,L3:2000kzo,LEPHiggsWorkingGroupforHiggsbosonsearches:2001ogs} and Tevatron \cite{D0:1999rfq,D0:2009hbc,D0:2009oou,D0:2009vff,CDF:2005acr,Yu:2009zzc,CDF:2009efz}. The current newest experimental data at the $13~ \mathrm{TeV}$ LHC and corresponding analyses are given in Refs. \cite{CMS:2019bfg,ATLAS:2018gfm}. In the high mass region, the decays of the charged Higgs boson to $t\bar{b}$, $c\bar{b}$ and $c\bar{s}$ have been studied at the LHC, reported by both ATLAS \cite{ATLAS:2018ntn,ATLAS:2021upq,ATLAS:2023bzb,Ivina:2022tfm} and CMS \cite{CMS:2020imj,CMS:2019rlz,CMS:2020osd,CMS:2018dzl}. The CMS collaboration has also analyzed the $13~ \mathrm{TeV}$ LHC dataset to search for the decays of $H^{\pm}$ to $W^{\pm}$ and other BSM bosons, such as $H^{\pm} \rightarrow W^{\pm} H^0$ \cite{CMS:2022jqc} and $H^{\pm} \rightarrow W^{\pm} A^0$ \cite{CMS:2019idx,Bhyun:2021chz}. These experimental measurements will be improved at the High-Luminosity LHC \cite{Cepeda:2019klc,deBlas:2019rxi} to operate in 2029. In addition, at future electron-positron colliders, more precise measurements on Higgs bosons can be performed with a cleaner environment and higher luminosity compared to the LHC. Several proposals for such colliders have been made: the International Linear Collider (ILC) \cite{Moortgat-Pick:2015lbx,Bambade:2019fyw}, the Future Circular Collider (FCC-ee) \cite{TLEPDesignStudyWorkingGroup:2013myl}, the Compact Linear Collider (CLIC) \cite{CLICPhysicsWorkingGroup:2004qvu,Aicheler:2012bya,Linssen:2012hp}, and the Circular Electron Positron Collider (CEPC) \cite{CEPCStudyGroup:2018ghi}.

\par
The charged Higgs boson pair production and its production associated with a $W$ boson at future $e^+e^-$ colliders have been widely studied. The leading order (LO) cross section and the next-to-leading order (NLO) electroweak (EW) corrections of $e^+e^- \rightarrow H^+H^-$ in the type-II THDM and the Minimal Supersymmetric Standard Model (MSSM) have been calculated in Refs.\cite{Diaz:1995kc,Arhrib:1998gr,Guasch:2001hk}, followed by the NLO calculations in the complex MSSM (cMSSM) \cite{Heinemeyer:2016wey} and the Inert Higgs Doublet Model (IHDM) \cite{Abouabid:2022rnd}. At the tree level, the $H^{\pm}W^{\mp}$ associated production at $e^+e^-$ colliders is highly suppressed by electron mass. The full one-loop contributions of $e^+e^-\rightarrow H^{\pm}W^{\mp}$ are given in Refs. \cite{Kanemura:1999tg,Arhrib:1999rg,Kanemura:2000si}, and the two-loop level corrections have been computed to meet the higher precision demand in the future \cite{Yang:2020msy}. Incidentally, studies on charged Higgs production at future muon colliders \cite{Han:2021udl} are also presented \cite{Akeroyd:1999xf,Hashemi:2013sja,Ouazghour:2023plc}.

\par
In the THDM, a discrete $Z_2$ symmetry is introduced to guarantee the absence of the tree-level Higgs-mediated flavor changing neutral currents (FCNCs). The four types of THDMs, referred to as type-I, type-II, type-X (lepton-specific) and type-Y (flipped), are distinguished by the representations of the $Z_2$ symmetry imposed \cite{Glashow:1976nt,Paschos:1976ay}. In this paper, we study the $e^+e^- \rightarrow H^+H^-$ process in the context of the THDM. We only focus on type-I and type-II THDMs, since there is no difference between type-I (type-II) and type-X (type-Y) THDMs on quark Yukawa couplings. The next-to-next-to-leading order (NNLO) $\text{QCD}\otimes\text{EW}$ corrections to $e^+e^- \rightarrow H^+H^-$ are calculated for the first time, and a discussion on numerical results are given, including the dependences on kinematic variables and THDM parameters. For completeness, we also compute the NLO EW corrections and the initial-state radiation (ISR) corrections to obtain the corrected cross section up to $O(\alpha\alpha_s)$.

\par
The rest of this paper is organized as follows. In Sec.\ref{sec:THDM}, we briefly introduce the THDM and recapitulate experimental and theoretical constraints on $m_{H^{\pm}}$ and $\tan\beta$. Descriptions of notations, techniques, and benchmark scenarios we adopt are provided in Sec.\ref{sec:description}. In Sec.\ref{sec:result}, our numerical results and corresponding analyses are presented, with some discussions on the $O(\alpha\alpha_s)$ corrections. Finally, a short summary is given in Sec.\ref{sec:summary}.

\section{THDM with $Z_2$ symmetry}
\label{sec:THDM}
\par
In contrast to the SM, the Higgs sector of the THDM comprises two complex scalar $SU(2)_L$ doublets $\Phi_i~ (i = 1, 2)$ with hypercharge $Y = +1$. In the scenario considered in this paper, a discrete $Z_2$ symmetry $\Phi_i \rightarrow (-1)^i \Phi_i~ (i = 1, 2)$ is introduced to restrict the Higgs sectors \cite{Glashow:1976nt,PhysRevD.15.1966}. Under this restriction, the most general renormalizable and $SU(2)_L \otimes U(1)_Y$ gauge invariant scalar potential has the form as
\begin{equation}
    \label{eq:potential}
    \begin{aligned}
        \mathcal{V}
    	& = m_{11}^{2} \Phi_{1}^{\dagger} \Phi_{1}+m_{22}^{2} \Phi_{2}^{\dagger} \Phi_{2}-m_{12}^{2} \left[\Phi_{1}^{\dagger} \Phi_{2}+\mathrm{h.c.}\right] +\frac{1}{2} \lambda_{1}\left(\Phi_{1}^{\dagger} \Phi_{1}\right)^{2}+\frac{1}{2} \lambda_{2}\left(\Phi_{2}^{\dagger} \Phi_{2}\right)^{2}\\
    	& +\lambda_{3}\left(\Phi_{1}^{\dagger} \Phi_{1}\right)\left(\Phi_{2}^{\dagger} \Phi_{2}\right)+\lambda_{4}\left(\Phi_{1}^{\dagger} \Phi_{2}\right)\left(\Phi_{2}^{\dagger} \Phi_{1}\right)+\frac{1}{2} \left[\lambda_{5}\left(\Phi_{1}^{\dagger} \Phi_{2}\right)^{2}+\mathrm{h.c.}\right].
    \end{aligned}
\end{equation}
The dimension-two term proportional to $m_{12}^2$ only softly breaks the $Z_2$ symmetry, and thus is tolerated. Due to the hermiticity of the scalar potential, the parameters $m^2_{11}, m^2_{22}, \lambda_{1}, \lambda_{2}, \lambda_{3}, \lambda_{4}$ must be real. Additionally, to ensure the tree-level $\mathcal{CP}$-conservation in the Higgs sector, the parameters $m^2_{12}$ and $\lambda_{5}$ are also set real \cite{Ferreira:2004yd,Gunion:2002zf,Gunion:2005ja}.\footnote{Other cases with $\mathcal{CP}$-violating terms are discussed in Refs.\cite{Branco:2011iw,Wu:1994ja}.} Consistent with the convention in Ref.\cite{Eriksson:2010zzb}, the two scalar doublets $\Phi_i~ (i= 1, 2)$ are parameterized as
\begin{equation}
    \Phi_{i}=\left(\begin{array}{c}
    \omega_{i}^{+} \\
\frac{1}{\sqrt{2}}\left(v_{i}+\xi_{i}+i \chi_{i}\right)
\end{array}\right),
\qquad
(i = 1, 2)
\end{equation}
with $\omega_{i}^{+}$ denoting the charged fields, $\xi_{i}$ the neutral $\mathcal{CP}$-even fields, and $\chi_{i}$ the neutral $\mathcal{CP}$-odd fields. The $v_{1,2}$ are the vacuum expectation values (VEVs) of the neutral components of $\Phi_{1,2}$, satisfying $v^{2}=v_{1}^{2} + v_{2}^{2}=(\sqrt{2}G_F)^{-1} \approx (246\ \mathrm{GeV})^{2}$. The two VEVs are real, required by the $\mathcal{CP}$-conservation, and always freely chosen to be positive. The mass eigenstates of the Higgs sector are obtained by the following linear transformations \cite{Altenkamp:2017ldc},
\begin{equation}
    \left(\begin{array}{c}{H^0} \\ {h^0}\end{array}\right)=
    \left(\begin{array}{cc}{\cos \alpha} & {\sin \alpha} \\ {-\sin \alpha} & {\cos \alpha}\end{array}\right)\left(\begin{array}{c}{\xi_{1}} \\ {\xi_{2}}\end{array}\right),
\end{equation}
\begin{equation}
    \left(\begin{array}{c}{G^{0}} \\ {A^0}\end{array}\right)=
    \left(\begin{array}{cc}{\cos \beta} & {\sin \beta} \\ {-\sin \beta} & {\cos \beta}\end{array}\right)\left(\begin{array}{c}{\chi_{1}} \\ {\chi_{2}}\end{array}\right),
\end{equation}
\begin{equation}
    \left(\begin{array}{c}{G^{\pm}} \\ {H^{\pm}}\end{array}\right)=
    \left(\begin{array}{cc}{\cos \beta} & {\sin \beta} \\ {-\sin \beta} & {\cos \beta}\end{array}\right)\left(\begin{array}{c}{\omega_{1}^{\pm}} \\ {\omega_{2}^{\pm}}\end{array}\right),
\end{equation}
where $\alpha$ represents the mixing angle of the neutral $\mathcal{CP}$-even Higgs sector, and the mixing angle $\beta$ for both charged and $\mathcal{CP}$-odd Higgs sectors relates to $v_{1,2}$ by $\tan \beta = v_{2}/v_{1}$. After the spontaneous symmetry breaking $SU(2)_L \otimes U(1)_Y \longrightarrow U(1)_{em}$, five physical Higgs bosons, $h^0$, $H^0$, $A^0$ and $H^{\pm}$, are obtained, with $G^{\pm}$ and $G^{0}$ absorbed by $W^{\pm}$ and $Z$ gauge bosons. The seven independent parameters of the THDM Higgs potential could be chosen as $m_{H^{\pm}}$, $m_{A^0}$, $m_{H^0}$, $m_{h^0}$, $\sin(\beta - \alpha)$, $\tan\beta$ and $\lambda_{5}$. The SM Higgs field is a linear combination of $h^0$ and $H^0$,
\begin{equation}
    \begin{aligned}
        h^{SM} & =  \xi_{1} \cos\beta + \xi_{2} \sin\beta           \\
               & =         h^0\sin(\beta-\alpha) + H^0\cos(\beta-\alpha).
    \end{aligned}
\end{equation}
To comply with SM physics, one of $h^0$ and $H^0$ in the THDM is SM-like \cite{Bernon:2015qea,Gunion:2002zf,Haber:1995be}. The two scenarios differ in the Yukawa couplings with a sign flip.
In this paper, the alignment limit, with the SM Yukawa structure, is applied where we suppose the SM Higgs boson discovered at the LHC to be the light neutral $\mathcal{CP}$-even Higgs boson $h^0$.

\par
To extend the $Z_2$ symmetry to the Yukawa sector, each quark and lepton doublet should be assigned $Z_2$ charge. The four types of the THDM correspond to the four distinct representations of $Z_2$ symmetry, which are listed in Table \ref{tab:types_of_2HDM}. As we can see from Table\ref{tab:types_of_2HDM}, yype-I (type-II) and type-X (type-Y) THDMs have the same Higgs-quark Yukawa couplings and different Higgs-lepton Yukawa couplings, while the latter are proportional to the masses of leptons and negligible in our work. Hence, in physical analyses, we mainly focus type-I and type-II THDMs and distinguish type-I (type-II) from type-X (type-Y) only when considering allowed parameter space. The Yukawa couplings of the charged Higgs boson to quarks can be expressed as
\begin{equation}
    \begin{aligned}
	    \mathcal{L}_{\text {Yukawa}}=&-\left[\frac{\sqrt{2} V_{\text{CKM}}}{v} \bar{u_i}\left(m_{u,i} \xi^{u,i} P_{L}+m_{d,i} \xi^{d,i} P_{R}\right) d_i H^{+}+ \text{h.c.}\right],
    \end{aligned}
    \label{eq:yukawa}
\end{equation}
where $V_{\text{CKM}}$ is the Cabibbo–Kobayashi–Maskawa (CKM) matrix, and $P_{L,R} \equiv (1 \mp \gamma_{5})/2$ are the chirality projection operators. $m_{u,i}$ and $m_{d,i}$ represent the up-type and down-type quark masses respectively. The coupling coefficients $\xi^{u,i}$ and $\xi^{d,i}$ are shown in Table\ref{tab:Yukawa_coupling}. For the Yukawa couplings between quarks and the charged Higgs boson, we only consider the third generation and neglect the others.
\begin{table*}
    \begin{tabular}{c|c|c|c|c}
    \hline \hline
	    & $u_{i}$ & $d_{i}$ & $e_i$ & $Z_{2}$ symmetry \\
    \hline
    type-I & $\Phi_{2}$ & $\Phi_{2}$ & $\Phi_{2}$  & $\Phi_{1} \rightarrow-\Phi_{1}$ \\
    type-II & $\Phi_{2}$ & $\Phi_{1}$ & $\Phi_{1}$ & $\left(\Phi_{1}, d_{i}, e_i \right) \rightarrow-\left(\Phi_{1}, d_{i}, e_i \right)$ \\
    type-X & $\Phi_{2}$ & $\Phi_{2}$ & $\Phi_{1}$ & $\left(\Phi_{1}, e_i \right) \rightarrow-\left(\Phi_{1}, e_i \right)$ \\
    type-Y & $\Phi_{2}$ & $\Phi_{1}$ & $\Phi_{2}$ & $\left(\Phi_{1}, d_{i}\right) \rightarrow-\left(\Phi_{1}, d_{i}\right)$ \\
    \hline \hline
    \end{tabular}
    \caption{Four types of THDMs and the corresponding representations of $Z_2$ symmetry. $u_i$, $d_i$ and $e_i~ (i = 1, 2, 3)$ represent the right-handed up-type quarks, down-type quarks and charged leptons, respectively.}
    \label{tab:types_of_2HDM}
\end{table*}
\begin{table*}
    \begin{tabular}{c|c|c|c|c}
        \hline \hline
        & type-I & type-II & type-X & type-Y \\
        \hline
        $\xi^{u,i}$ & $\cot \beta$ & $\cot \beta$ & $\cot \beta$ & $\cot \beta$ \\
        $\xi^{d,i}$ & $\cot \beta$ & $-\tan \beta$ & $\cot \beta$ & $-\tan \beta$ \\
        \hline \hline
    \end{tabular}
    \caption{Yukawa couplings of $H^{\pm}$ to quarks in different types of THDMs.}
    \label{tab:Yukawa_coupling}
\end{table*}

\par
Constraints on THDM parameters from experimental data have been analyzed detailedly in Ref.\cite{Haller:2018nnx}.
In the following, we replenish relevant recent studies and specifically explain the constraints on $m_{H^{\pm}}$ and $\tan\beta$ that the $O(\alpha\alpha_s)$ corrections depend on.
\begin{description}
	\item[	1) Direct searches for BSM Higgs bosons] Constraints according to data from the $13\ \mathrm{TeV}$ LHC with $36\ \mathrm{fb}^{-1}$ are evaluated in Ref.\cite{Aiko:2020ksl}. In the alignment limit, the calculation results require $\tan\beta\gtrsim 2$ for type-I, $2.3\lesssim\tan\beta\lesssim8$ for type-II, $\tan\beta\gtrsim 8$ for type-X and $\tan\beta\gtrsim 1.2$ for type-Y. The direct searches at the LEP give the lower bound of $m_{H^{\pm}}$ as $72.5\ \mathrm{GeV}$ for type-I and $80\ \mathrm{GeV}$ for type-II \cite{ALEPH:2013htx}. 
	\item[	2) Flavor constraints] The B meson flavor violating decay $B\rightarrow X_s\gamma$ gives strict constraints as $m_{H^{\pm}} \gtrsim 800\ \mathrm{GeV}$ for type-II and type-Y when $\tan\beta\gtrsim1$ \cite{Misiak:2020vlo}. For $\tan\beta$, the upper bound is $\tan\beta\lesssim 20$ at $m_{H^{\pm}} = 800\ \mathrm{GeV}$ for type-II, constrained by the measurements of the branching ratio $\mathcal{B}(B_s\rightarrow\mu\mu)$ at LHCb. The anlysis on $\mathcal{B}(B_d\rightarrow\mu\mu)$ provides the lower bound $\tan\beta\gtrsim 3$ at $m_{H^{\pm}}=120\ \mathrm{GeV}$ for type-I and type-X, which gradually decreases to $\tan\beta\gtrsim 1.2$ with $m_{H^{\pm}}$ increasing to $780\ \mathrm{GeV}$ \cite{CMS:2014xfa,LHCb:2017rmj}.
	\item[	3) Anomalous magnetic moment of the muon] THDM predictions up to the two-loop level of the anomalous magnetic moment of the muon are evaluated in Ref.\cite{Cherchiglia:2016eui}, and allowed regions on the $\tan\beta-m_{H^{\pm}}$ plane for four types of THDMs are given in Ref.\cite{Haller:2018nnx}. The region $\tan\beta\lesssim 2$ is excluded for all types of THDMs at the $95\%$ confidence level, and the limit becomes stricter with larger $m_{H^{\pm}}$. 
	\item[  4) SM Higgs boson coupling measurements] The signal strength measurements of $h$ from the LHC provide constraints mainly on $\beta\pm\alpha$ \cite{ATLAS:2015ciy}. For type-II, type-X, and type-Y, the allowed parameter space is restricted in two narrow bands where $\cos(\beta+\alpha)$ or $\cos(\beta-\alpha)$ are highly limited close to $0$ separately, and they become even narrower with increasing $\tan\beta$. In contrast, the constraints on $\alpha$ and $\beta$ are looser for type-I, allowing the region $-0.25\lesssim \cos(\beta-\alpha)\lesssim 0.26$.
\end{description}
THDM parameters also have the following theoretical and experimental constraints, giving inexplicit bounds on $m_{H^{\pm}}$ and $\tan\beta$. Within the region of $m_{H^{\pm}} \in [130,\ 1500]\ \mathrm{GeV}$ and $\tan\beta \in [2,\ 50]$, the following constraints are checked for four types of THDMs in the benchmark scenario to be mentioned in subsection\ref{subsec:NLO} with the package \texttt{2HDMC} \cite{Eriksson:2009ws}. 
\begin{description}
        \item[  5) Oblique parameters] The EW oblique parameters S, T and U, with the THDM contribution \cite{Grimus:2007if,Grimus:2008nb} included, should stay in the range allowed by experimental results, whose recent values given by Ref.\cite{ParticleDataGroup:2022pth} are $S=-0.02\pm0.10$, $T=0.03\pm0.12$ and $U=0.01\pm0.11$.
	\item[  6) Vacuum stability] The THDM Higgs potential must be positive definite to ensure vacuum stability, with necessary and sufficient conditions given in \cite{Nie:1998yn}. We also consider further studies on the metastability of the THDM vacuum, requiring a so called true vacuum condition written as $m_{12}^2(m_{11}^2-k^2m_{22}^2)(\tan\beta-k)>0$ with $k=\sqrt[4]{\lambda_1/\lambda_2}$ \cite{Barroso:2013awa,Branchina:2018qlf}.
	\item[  7) Perturbative unitarity] For $2\rightarrow 2$ processes of Higgs bosons and longitudinal parts of EW gauge bosons, eigenvalues of S-wave amplitude matrices are constrained in the form of $|\mathrm{Re}(x_i)|<\xi$, where $\xi$ is a constant determined by specific forms of the matrices. This constraint is demanded by the tree-level perturbative unitarity. Otherwise, tree-level calculations are unreliable and meaningless. \cite{Kanemura:2015ska}. 			
\end{description}

\section{Description of calculation}
\label{sec:description}
\subsection{Conventions and setup}
\label{sec:setup}
\par
We consider the charged Higgs boson pair production on the electron-positron collider
\begin{equation}
        e^+(p_1)+e^-(p_2) \rightarrow H^+(p_3) + H^-(p_4).
\end{equation}
Neglecting the electron mass, the momenta in the center-of-mass frame can be written as
\begin{equation}
        p_{1,2}= E(1,0,0,\pm 1), \quad
        p_{3,4}= E(1,\pm\kappa \sin\theta,0, \pm\kappa \cos\theta),
\end{equation}
where $E$ denotes the beam energy, $\theta$ is the scattering angle between $e^+$ and $H^+$, and $m_{H^\pm}$ is the mass of the charged Higgs boson, with $\kappa \equiv \sqrt{1-4m_{H^\pm}^2/s}$. The Mandelstam variables are
\begin{equation}
        \begin{aligned}
                s &= (p_1+p_2)^2 = 4E^2, \\
                t &= (p_1-p_3)^2 = m_{H^\pm}^2-2E^2+2E^2\kappa \cos\theta, \\
                u &= (p_1-p_4)^2 = m_{H^\pm}^2-2E^2-2E^2\kappa \cos\theta.
        \end{aligned}
\end{equation}

\par
Numerical results in two input parameter schemes are compared in our work: $\alpha(0)$ and $\alpha(m_Z)$, and the latter is applied by default. In the $\alpha(0)$ scheme, the EW coupling constant $\alpha$ is input as the fine-structure constant $\alpha(0)$, with the charge renormalization constant given by
\begin{equation}
	\delta Z_e|_{\alpha(0)} = \frac{1}{2}\Pi^{\gamma\gamma}(0)-\frac{s_W}{c_W}\frac{\Sigma_T^{\gamma Z}(0)}{m_Z^2}.
\end{equation}
In the $\alpha(m_Z)$ scheme, $\alpha$ is evolved, using renormalization group equations, from the Thomson limit to the $m_Z$ pole and written as $\alpha(m_Z)=\alpha(0)/(1-\Delta\alpha)$, with
\begin{equation}
	\delta Z_e|_{\alpha(m_Z)} = \delta Z_e|_{\alpha(0)} -\frac{1}{2}\Delta\alpha,
\end{equation}
where $\Delta\alpha$ is defined as
\begin{equation}
	\Delta\alpha = \Pi^{\gamma\gamma}_{f\neq t}(0)
		   -\mathrm{Re} \Pi^{\gamma\gamma}_{f\neq t}(m_Z^2)
		   ,\quad \Pi^{\gamma\gamma}(Q^2) \equiv
		   \frac{\Sigma_T^{\gamma \gamma}(Q^2)}{Q^2}.
\end{equation}
When $Q=0$, nonperturbative strong coupling effects from QCD become nonnegligible, the hadronic contributions should be absorbed into a nonperturbative parameter $\Delta\alpha^{(5)}_{\mathrm{had}}$, fixed by the experiment measurement. Then $\Delta\alpha$ is determined via
\begin{equation}
	\Delta\alpha = \Delta\alpha^{(5)}_{\mathrm{had}} + \Delta\alpha_{\mathrm{lep}},
\end{equation}
and the leptonic contributions $\Delta\alpha_{\mathrm{lep}}$ are nonzero at $O(\alpha)$ but zero at $O(\alpha\alpha_s)$.

\par
We generate Feynman diagrams and amplitudes with the \texttt{FeynArts} \cite{Hahn:2000kx} package, using the Feynman rules of the THDM in Ref.\cite{Altenkamp:2017ldc}. To obtain UV finite results, we adopt the on-shell renormalization scheme. The renormalization constant $\delta Z_{H^{\pm}}$ is given by
\begin{equation}
	\delta Z_{H^{\pm}} = -\mathrm{Re}\frac{\partial\Sigma^{H^{\pm}}(k^2)}{\partial k^2}\Big|_{k^2=m^2_{H^{\pm}}},
\end{equation}
while others are already defined in the SM renormalization and can be found in Ref.\cite{Denner:1991kt}.

\par
At the tree-level, two Feynman diagrams, mediated by the $\gamma$ and Z $s$-channel exchange, contribute to the $e^+e^- \rightarrow H^+H^-$ process. Feynman diagrams involving electron-Higgs coupling vertices are neglected, and diagrams of one-loop and two-loop neglected for the same reason are not discussed repeatedly in the following. The differential cross section is 
\begin{equation}
	(\frac{d\sigma}{d\Omega})_0 = \frac{\alpha^2\kappa^3}{8s}\left(1+g_H^2\frac{g_V^2+g_A^2}{(1-m_Z^2/s)^2}-\frac{2g_Hg_V}{1-m_Z^2/s}\right)\sin^2\theta,
\end{equation}
with $g_V = (1-4s_W^2)/(4c_W s_W)$, $g_A = 1/(4c_W s_W)$, $g_H = (s_W^2-c_W^2)/(2c_W s_W)$, $c_W \equiv \cos\theta_W$, $s_W \equiv \sin\theta_W$.

\subsection{NLO EW corrections}
\label{subsec:NLO}
\par
The NLO EW correction, or equivalently the $O(\alpha)$ correction, is contributed by one-loop and real radiation diagrams. Dimensional regularization is adopted to regularize UV singularities. In real radiation contributions, the IR divergence comes from the phase space integration. The IR divergence from virtual corrections is presented by diagrams where a virtual photon is exchanged in loops. We address the infrared (IR) singularities by applying photon mass regularization and the two cutoff phase space slicing methods \cite{Dittmaier:1999mb,Denner:2000bj}. A small energy cutoff $\Delta E = \delta_s\sqrt{s}/2$ is introduced for the separation of soft and hard real radiation photon energy. Then the IR divergence is included in the soft region and will cancel with the counterpart in the virtual contribution. The asymptotic expression of the soft bremsstrahlung is studied in Ref.\cite{Yennie:1961ad}, which is the product of the Born cross section and the soft factor in the soft-photon limit $k\rightarrow 0$
\begin{equation}
\label{eq:SB}
	d\sigma_{\mathrm{S}} = -d\sigma_0 \frac{\alpha(0)}{2\pi^2}
	\int_{|{\bf k}|\le\Delta E}\frac{d^3{\bf k}}{2E_k}
	\sum_{i,j=1}^nQ_iQ_j\frac{p_i\cdot p_j}{p_i\cdot k  p_j\cdot k},
\end{equation}
where $k$ is the momentum of the soft photon, $p_i$ and $Q_i$ denote the momentum and charge flowing into the diagrams with the $i$-th external particle respectively, and $n$ is the number of external particles except the soft photon. The specific expression of Eq.~\eqref{eq:SB} for this process is given in Ref.\cite{Arhrib:1998gr}. Considering the hard region, the external particles are all massive. Hence, no collinear IR divergence should emerge in principle. However, the tiny electron mass induces a quasi-collinear IR divergence from the initial state radiation. Therefore, we continue to separate the hard part, using an angle cutoff $\Delta\theta$, into collinear and noncollinear regions. The cross section of the hard collinear part is factorized as a convolution of the Born cross section and the collinear factor
\begin{equation}
	d\sigma_{\mathrm{HC}} = \sum\limits_{i=1,2}\frac{\alpha}{2\pi}
	 \int_{0}^{1-\delta_s}dx_id\sigma_0(x_ip_i)
	 \left\{\frac{1+x^2_i}{1-x^2_i}\left[\ln\left(\frac{s}{m^2_e}\frac{\delta_c}{2}\right)-1\right]+(1-x_i)\right\},
\end{equation}
where $x_i$ represents the fractions of the initial state particle momenta left after the collinear bremsstrahlung. The remaining hard noncollinear region can be computed directly with the Monte Carlo method. In addition, we add the ISR correction with a leading logarithmic approximation \cite{Beenakker:1996kt,Denner:2000bj} (Although it contains higher order corrections, we include it in the NLO correction for brevity in discussions). Summing these terms, the total $O(\alpha)$ correction is composed of five parts
\begin{equation}
	\sigma_{\mathrm{NLO}} = \sigma_{\mathrm{V}}(\Delta_{\mathrm{UV}},\lambda) + \sigma_{\mathrm{S}}(\lambda,\Delta E) +\sigma_{\mathrm{HC}}(\Delta E,\Delta\theta) + \sigma_{\mathrm{H\bar{C}}}(\Delta E,\Delta\theta) + \sigma_{\mathrm{ISR}},
\end{equation}
where $\sigma_{\mathrm{V}}$, $\sigma_{\mathrm{S}}$, $\sigma_{\mathrm{HC}}$, $\sigma_{\mathrm{H\bar{C}}}$ and $\sigma_{\mathrm{ISR}}$ represent the virtual, soft, hard collinear, hard noncollinear and ISR contributions respectively.

\par
Concerning the benchmark scenario, the following rules are applied to constrain THDM parameters that the NLO correction depends on. We adopt the alignment limit by fixing $\sin(\beta-\alpha) = 1$. As the THDM potential shows, $\lambda_5$ modifies Higgs self couplings and is fixed to $0$ because we do not concentrate on this topic in our work. Furthermore, $m_H = m_A = m_{H^{\pm}}$ is set so that perturbative unitarity is always ensured in our parameter region. 

\par
At the $O(\alpha)$, we perform the evaluation with modified \texttt{FormCalc} and \texttt{LoopTools} packages \cite{Hahn:1998yk,Hahn:2006qw,Hahn:1999mt,Hahn:2010zi}. We numerically check the independence of the total NLO correction with two unphysical cutoffs $\delta_s$ and $\Delta\theta$, ranging from $10^{-6}$ to $10^{-4}$, and the fictitious photon mass $\lambda$, ranging from $10^{-10}$ to $1$. The cancellation of UV divergences is also tested with an extensive range variation of the dimensional regularization parameter $\Delta_{\mathrm{UV}}$. Our results are consistent with \cite{Guasch:2001hk}, and we also check with the fermion part contribution tabulated in \cite{Arhrib:1998gr}.

\subsection{NNLO QCD$\otimes$EW corrections}
\par
\label{subsec:NNLO}
\begin{figure*}
    \centering
    \includegraphics[scale=1.1]{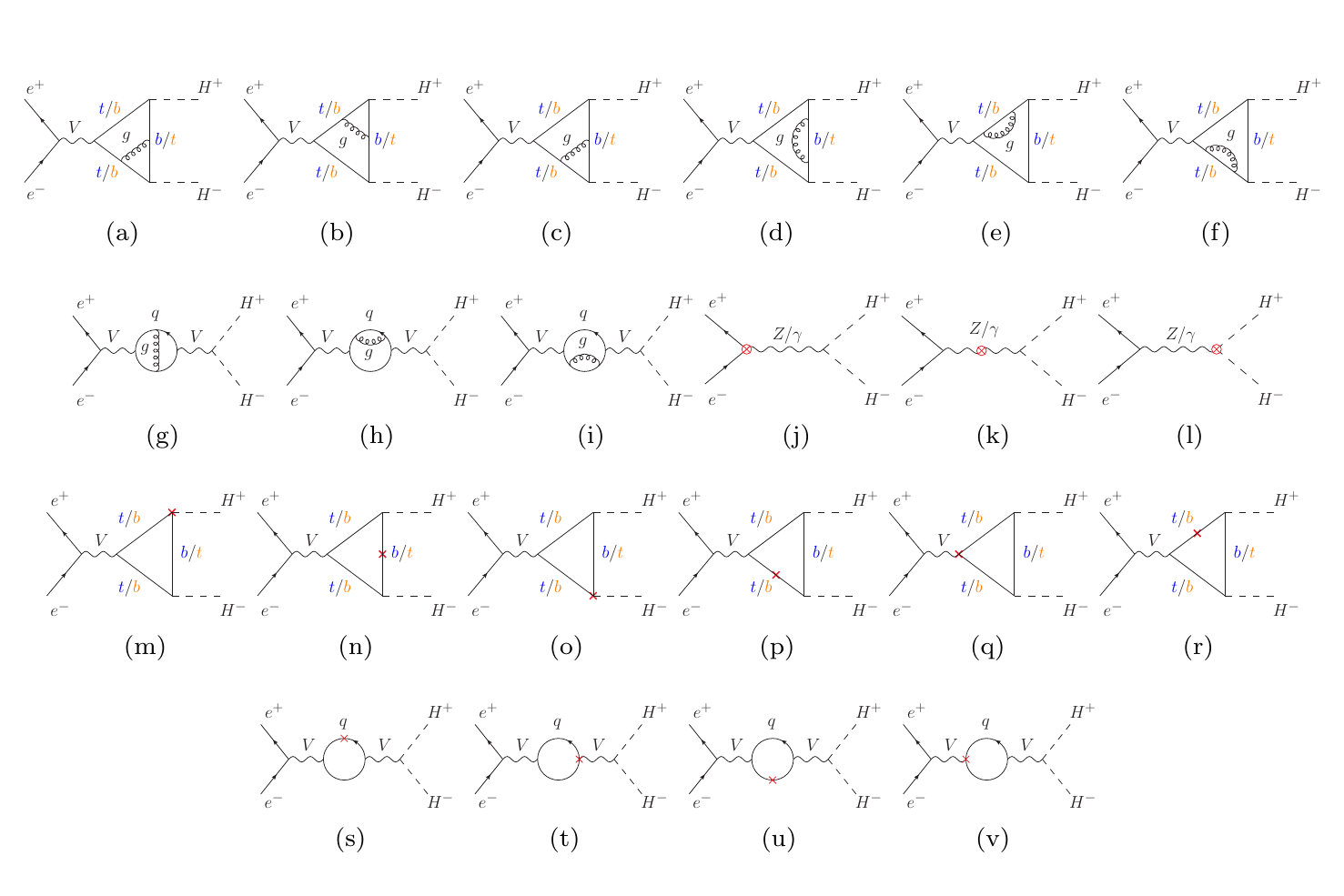}
    \caption{Feynman diagrams for the $e^+e^-\rightarrow H^+H^-$ process at NNLO. $V$ can be $\gamma$ or $Z$. $t$ and $b$ quarks in blue denote one diagram ,and in orange denote the other. The red cross represents the renormalization constants at the $O(\alpha_s)$, while the red circle cross represents that at $O(\alpha\alpha_s)$. }
    \label{fig:feynman_nnlo}
\end{figure*}
At NNLO, which is also the $O(\alpha\alpha_s)$, all two-loop diagrams can be generated by adding a gluon line to the quark-loop diagrams at the $O(\alpha)$. In the counterterm diagrams shown in Fig.\ref{fig:feynman_nnlo}, the red cross represents the renormalization constants at $O(\alpha_s)$, while the circle cross represents that at the $O(\alpha\alpha_s)$. It is noticeable that the $O(\alpha_s)$ quark wave function counterterms are canceled by virtue of the QED-like Ward identity. Then the $O(\alpha_s)$ counterterms have only contributions from quark mass renormalization constants written as \cite{Bernreuther:2004ih}
\begin{equation}
    \delta_{m_{q}} = -m_{q}\frac{\alpha_{s}}{3\pi}C(\epsilon)\left(\frac{4\pi\mu^{2}}{m_{q}^{2}}\right)^{\epsilon}\frac{(3-2\epsilon)}{\epsilon(1-2\epsilon)}\Gamma(1+\epsilon),
\end{equation}
where $\mu$ is the energy scale and $\epsilon$ is the dimensional regular. The diagrams with $s$-channel Vector-Scalar mixing are not shown because they can be analytically proven to make no contribution using the Dirac equation.

\par
We evaluate Dirac traces and perform tensor decomposition with the Passarino-Veltman method using the \texttt{FeynCalc} package \cite{Mertig:1990an,Shtabovenko:2016sxi}. Form factors independent of loop momenta are then extracted from amplitudes with remaining scalar integrals in the form called the Feynman integral 
\begin{equation}
    \mathcal{I}(\left\{a_{1}\ldots a_{n} \right\},d,\eta) = \int \prod_{j=1}^{L}d^{d}l_{j}\prod_{k=1}^{n}\frac{1}{(D_{k}+i\eta)^{a_{k}}},
\end{equation}
where $d = 4-2\epsilon$ and $D_{k} \equiv q_{k}^{2}-m_{k}^2$ ($q_k$ are linear combinations of loop momenta $l_j$ and external momenta $p_i$) are propagators belonging to the same family. $\eta$ is an infinitesimal auxiliary mass and $L$ is the number of loops. For each family, we reduce the scalar integrals into combinations of a smaller set of integrals called master integrals (MIs) with the integration-by-parts (IBP) method \cite{Tkachov:1981wb, Chetyrkin:1981qh}, utilizing the \texttt{KIRA} package \cite{Maierhofer:2017gsa} which adopts Laporta's algorithm \cite{Laporta:2000dsw}.

\par
We numerically compute the MIs using the differential equation (DE) method mentioned in Refs.\cite{Liu:2017jxz,Ahnert:2011}. For a given Feynman integral family, derivatives of an integral to kinematics could be expressed as linear combinations of MIs by using the IBP identities. Thus, we are able to construct a linear differential equation system as
\begin{equation}
	\frac{\partial\vec{\mathcal{I}}(x)}{\partial x} = \mathcal{M}(x).\vec{\mathcal{I}}(x),
\end{equation}
where $\vec{\mathcal{I}}$ represents a complete set of MIs of the given family, and the derivative variable $x$ can be Mandelstam variables, particle masses, or the auxiliary mass $\eta$.

\par
In this work, we utilize our private codes to evaluate MIs in which the derivative variable is chosen as the auxiliary mass $\eta$. The boundary condition of DEs is determined at $\eta \rightarrow \infty$ where $\eta$ in MIs is extracted as factors and the remainders are simple vacuum integrals analytically evaluated to three-loop order in Refs.\cite{Davydychev:1992mt,Broadhurst:1998rz,Kniehl:2017ikj}. We evolve DEs from a point $\eta_i$ to another point $\eta_j$ successively with the help of the \texttt{odeint} package\cite{Ahnert:2011}, thus realizing the flow of $\eta$ from the boundary to the physical value point. The coefficient matrix of DEs is transformed into the normalized Fuchsian form with the \texttt{epsilon} package\cite{Prausa:2017ltv}, so that Feynman integrals could be expanded to asymptotic series and matched at a point near $\eta=0$ to determine unknown coefficients. The physical results of Feynman integrals are finally obtained at the limitation of $\eta \rightarrow 0^+$
\begin{equation}
        \mathcal{\vec{I}}(\left\{a_{1}\ldots a_{n} \right\},d,0) = \lim_{\eta\rightarrow 0^{+}} \mathcal{\vec{I}}(\left\{a_{1}\ldots a_{n} \right\},d,\eta).
\end{equation}
To ensure reliability, we also check the MIs numerically with a package called \texttt{AMFlow} \cite{Liu:2022chg}. After all the contributions at $O(\alpha\alpha_s)$ are summed, we have numerically checked the cross section is UV-finite and the precision of our results is higher than 30 digits, sufficient for the $O(\alpha\alpha_s)$ correction. 

\par
Two-loop Feynman integrals from diagrams of the self energy correction topology, drawn in Fig.\ref{fig:feynman_nnlo}(g-i), can be divided into three families $\mathcal{F}_i$ $(i=1,2,3)$. With $m_q$ denoting $m_t,m_b$ and $0$ for $\mathcal{F}_{1,2,3}$ respectively, the propagators read
\begin{align*}
   &D_1 = l_1^2 - m_q^2 ,\quad
    D_2 = l_2^2 - m_q^2 ,\quad
    D_3 = (l_1-l_2)^2 , \\
   &D_4 = (p_3+p_4+l_1)^2 - m_q^2 , \quad
    D_5 = (p_3+p_4+l_2)^2 - m_q^2 ,\quad
\end{align*}
where propagators with $q=u,d,c,s$ belong to $\mathcal{F}_3$ because the masses of these quarks are all neglected. Feynman integrals from the diagrams drawn in Fig.\ref{fig:feynman_nnlo}(a-f) belong to $\mathcal{F}_4$ and $\mathcal{F}_5$. These diagrams have the $VH^+H^-$ vertex correction topology and show quarks marked in blue for $\mathcal{F}_4$ and orange for $\mathcal{F}_5$. Corresponding propagators of $\mathcal{F}_4$ are chosen as
\begin{align*}
   &D_1 = l_1^2 - m_b^2 ,\quad
    D_2 = l_2^2 - m_b^2 ,\quad
    D_3 = (l_1-l_2)^2,\quad
    D_4 = (l_1+p_3)^2-m_t^2 , \\
   &D_5 = (l_2+p_3)^2-m_t^2 , \quad
    D_6 = (l_1-p_4)^2 - m_t^2 ,\quad
    D_7 = (l_2-p_4)^2 - m_t^2 ,
\end{align*}
while that of $\mathcal{F}_5$ can be obtained by exchanging top and bottom quarks in $\mathcal{F}_4$.

\section{Numerical results and discussion}
\label{sec:result}
\par
The following physical parameters are adopted in this section \cite{ParticleDataGroup:2022pth,Davier:2019can}
\begin{align*}
	&\alpha(0) = 1/137.035999084, && \Delta\alpha_{\mathrm{had}}^{(5)} = 0.02753, && \alpha(m_Z) = 1/128.947, \\
	&\alpha_s(m_Z) = 0.1179, && m_Z = 91.1876\ \mathrm{GeV}, && m_W = 80.377\ \mathrm{GeV}, \\
	&m_h = 125.25\ \mathrm{GeV}, && m_t = 172.5\ \mathrm{GeV}, && m_b = 4.78\ \mathrm{GeV}.
\label{eq:inputs}
\stepcounter{equation}\tag{\theequation}
\end{align*}
Masses of all other fermions are neglected whenever possible. We utilize the \texttt{RunDec} package \cite{Chetyrkin:2000yt,Herren:2017osy} to evolve the strong coupling constant $\alpha_s(\mu)$ to the scale of the colliding energy $\mu = \sqrt{s}$ from scale $\mu = m_Z$. 

\par
\begin{figure*}
    \centering
    \includegraphics[scale=0.78]{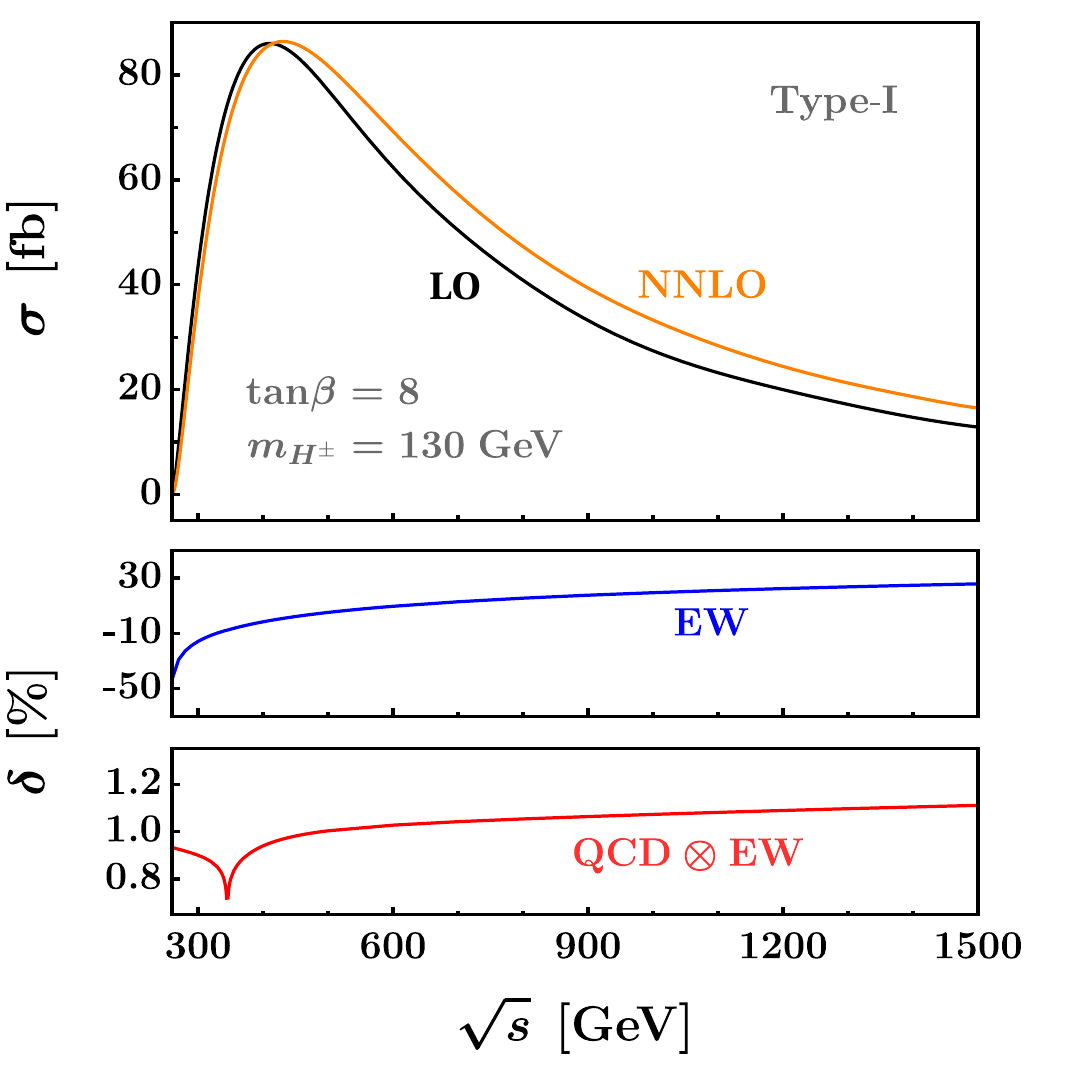}
    \includegraphics[scale=0.78,trim= 40pt 0 0 0]{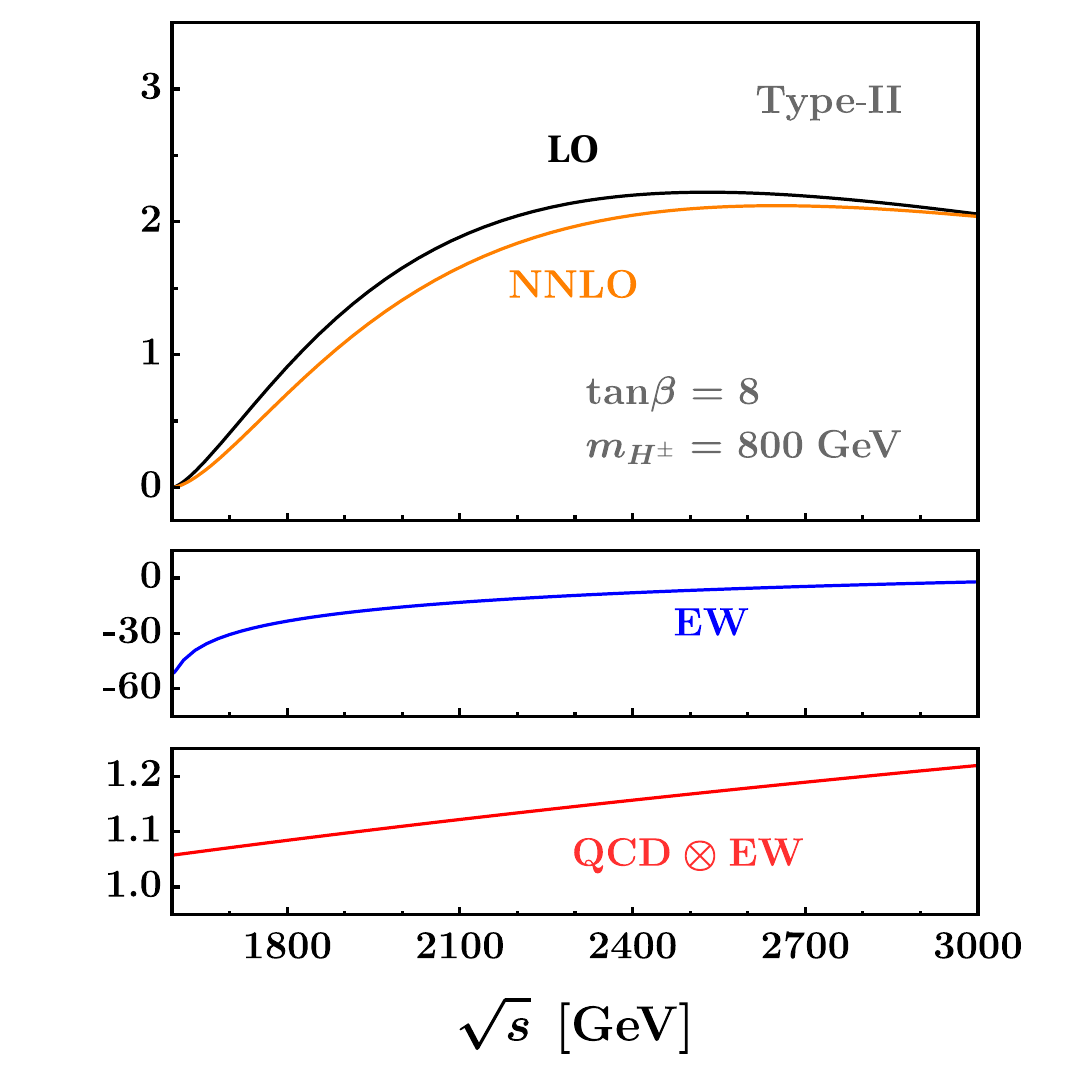}
	\caption{Corrected cross sections of $e^+e^-\rightarrow H^+H^-$ up to NNLO in association with Born cross sections and corresponding relative corrections as functions of the colliding energy.}
    \label{fig:ss}
\end{figure*}
We present the numerical results in both type-I and type-II THDMs as functions of $\sqrt{s}$ in Fig.\ref{fig:ss}. As shown in the top panel, the behaviors of the $O(\alpha\alpha_s)$ integrated cross sections are pretty similar to those of the Born. For the type-I THDM, the Born cross section reaches its maximum of $85.62$ fb at $\sqrt{s}\approx 409 \ \mathrm{GeV}$, while the corrected cross section up to NNLO reaches a maximum of $86.35$ fb at $\sqrt{s}\approx 430 \ \mathrm{GeV}$. The $O(\alpha)$ relative correction increases from $-41.86\%$ to $25.72\%$ with the increment of $\sqrt{s}$ from $261 \ \mathrm{GeV}$ to $1500 \ \mathrm{GeV}$. The $O(\alpha\alpha_s)$ relative correction increases slowly from $0.93\%$ at $\sqrt{s}=261 \ \mathrm{GeV}$ to $1.11\%$ at $1500 \ \mathrm{GeV}$. Notice that a resonance peak appears at $\sqrt{s}=2m_t=345 \ \mathrm{GeV}$, induced by the $t$ quark loop integrals at the threshold. For the type-II THDM, the cross sections are small compared to that of the type-I THDM, no more than $2.23\ \mathrm{fb}$, due to the large charged Higgs boson mass. The $O(\alpha)$ relative correction increases from $-46.18\%$ to $-2.12\%$ with the increment of $\sqrt{s}$ from $1601 \ \mathrm{GeV}$ to $3000 \ \mathrm{GeV}$, and the $O(\alpha\alpha_s)$ relative correction increases from $1.06\%$ to $1.22\%$. 

\par
\begin{figure*}
    \centering
    \includegraphics[scale=0.78]{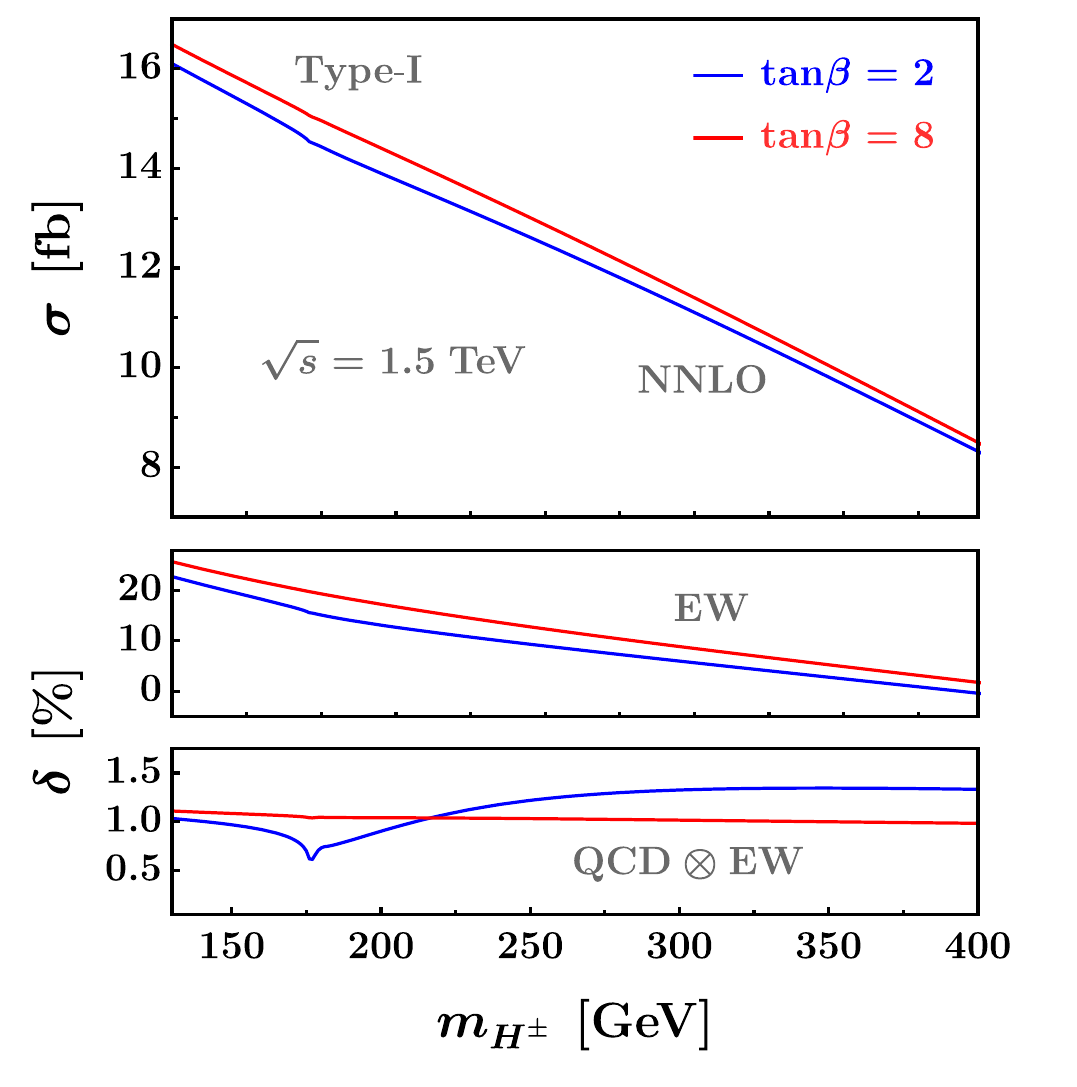}
    \includegraphics[scale=0.78,trim= 40pt 0 0 0]{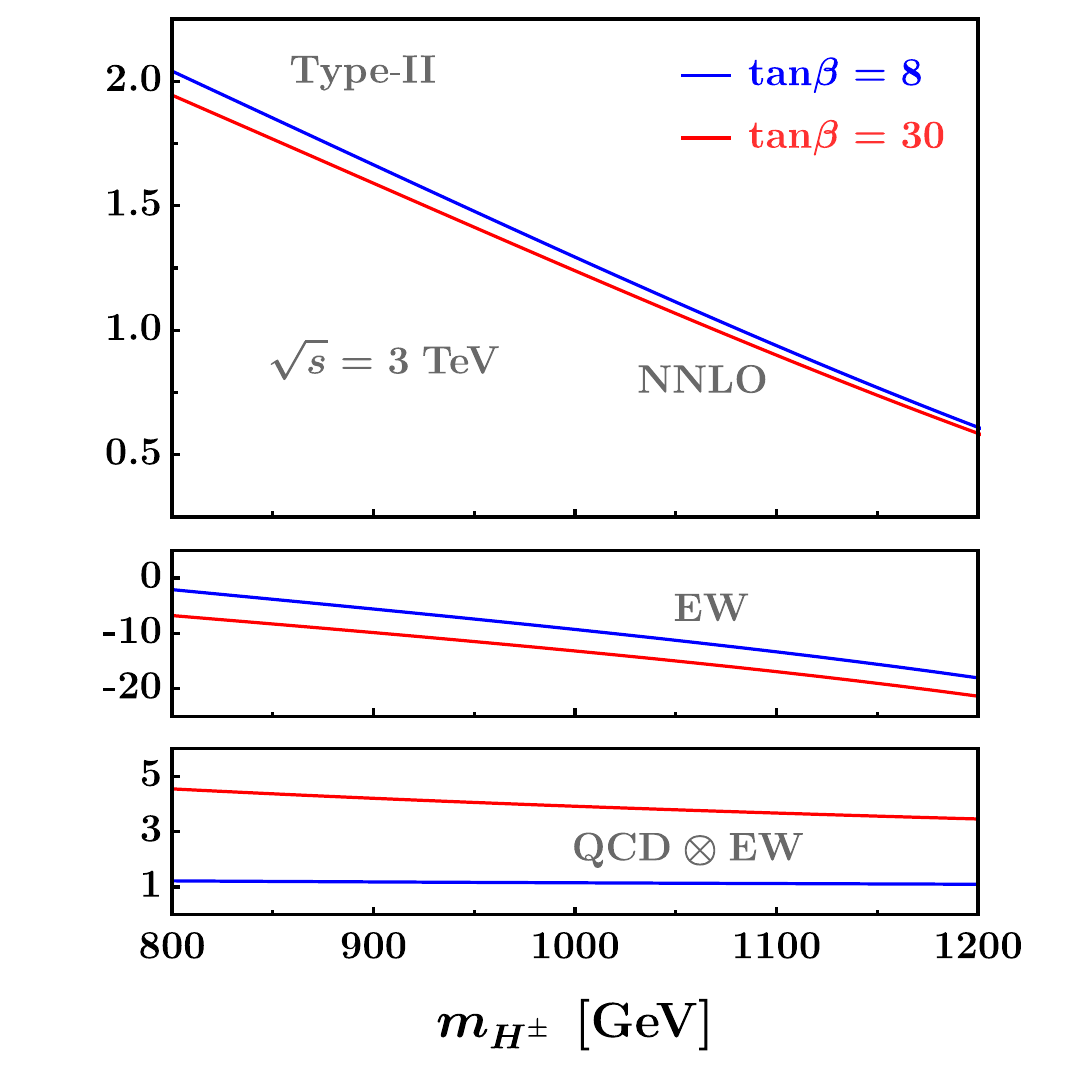}
	\caption{Corrected cross sections of $e^+e^-\rightarrow H^+H^-$ up to NNLO and the corresponding relative corrections as functions of $m_{H^{\pm}}$.}
    \label{fig:mhp}
\end{figure*}
Drawn in Fig.\ref{fig:mhp} are the $O(\alpha\alpha_s)$ results at different values of $\tan\beta$ as functions of the charged Higgs boson mass. As shown in the top panel, the cross sections decrease equably with the charged Higgs boson mass increment. For type-I, we can see from the middle and bottom panels that the $O(\alpha)$ relative corrections can be more significant than $20\%$ while the $O(\alpha\alpha_s)$ relative corrections are near $1\%$. There is a peak in $O(\alpha)$ and $O(\alpha\alpha_s)$ corrections (although not obvious in the former) at $m_{H^\pm} = m_t + m_b = 177.28 \ \mathrm{GeV}$, owing to the resonance effect. For type-II, the $O(\alpha\alpha_s)$ relative corrections are larger and reach $4.54\%$ at $\tan\beta=30$ and $m_{H^{\pm}}=800\ \mathrm{GeV}$. By analyzing Eq.~\eqref{eq:yukawa}, notice that the Yukawa coupling vertex is proportional to $\tan\beta$ in the type-I THDM. Thus the $O(\alpha\alpha_s)$ correction as a function of $\tan\beta$ can be expressed in the form of $A_1+B_1 \cot^2\beta$, where $A_1$ is the contribution from diagrams of the self energy correction topology and $B_1\cot^2\beta$ comes from the $VH^+H^-$ vertex correction diagrams. The curves in the bottom left panel of different $\tan\beta$ intersect at the same point for this reason, at which $B_1=0$ exactly. Also from the analysis of the Yukawa coupling vertex, such an intersection does not necessarily appear in the type-II THDM, where the $\beta$ dependence is in the form of $A_2+B_2 \cot^2\beta +C_2\tan^2\beta$. The above conclusions about $\tan\beta$ dependence are also valid at $O(\alpha)$ with similar analyses, although the crosspoint does not necessarily appear in the region we plot.

\par
\begin{figure*}
    \centering
    \includegraphics[scale=0.78]{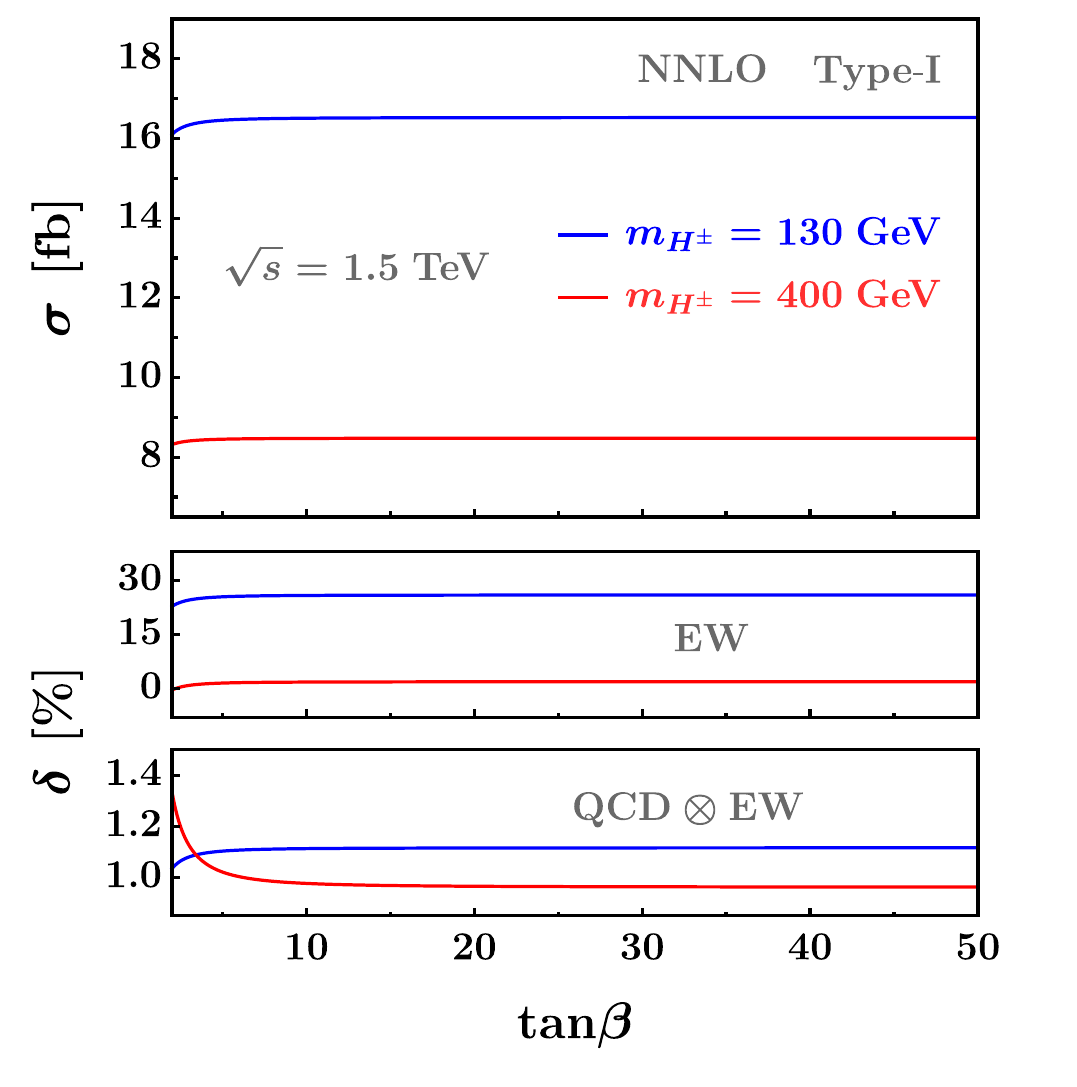}
    \includegraphics[scale=0.78,trim= 40pt 0 0 0]{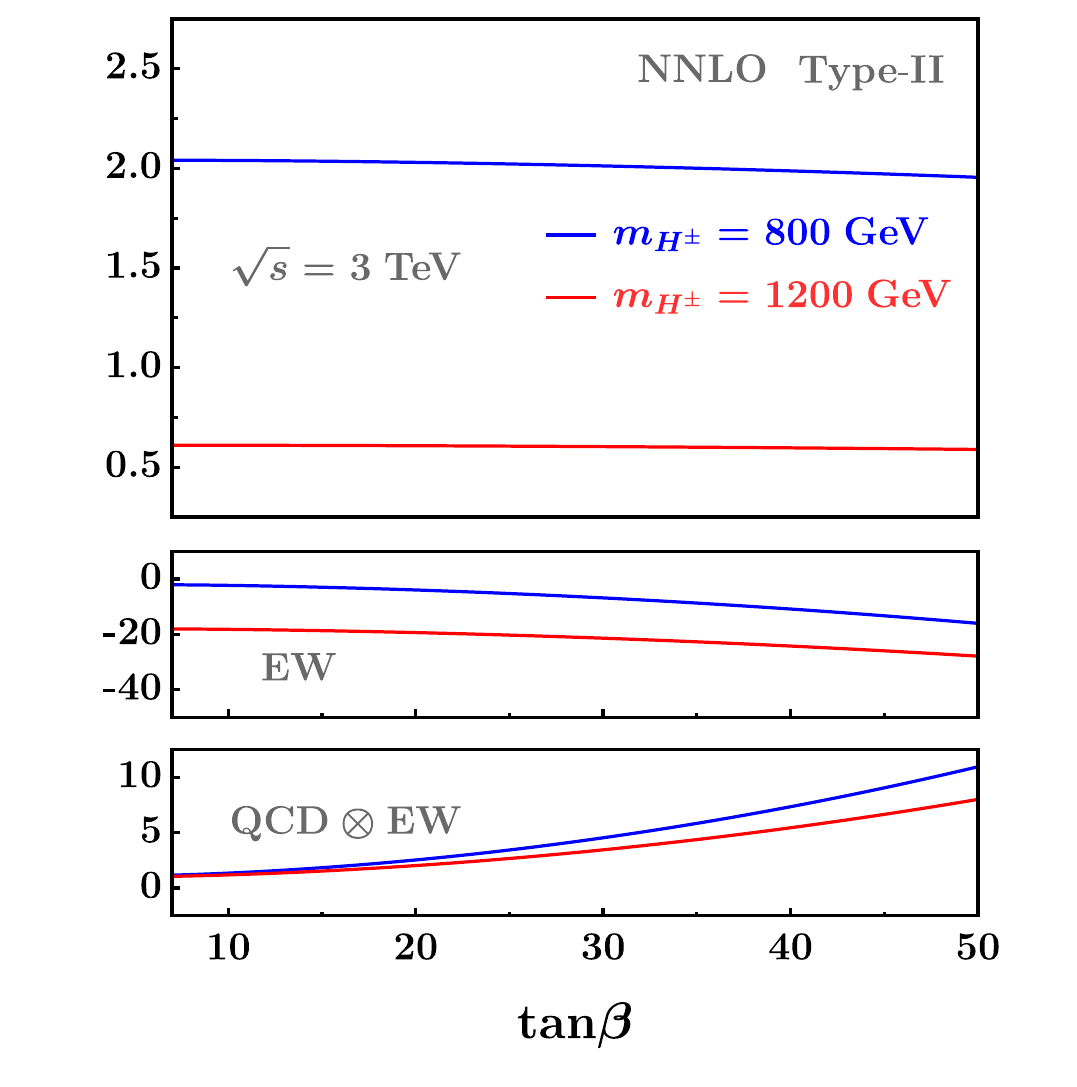}
	\caption{Corrected cross sections of $e^+e^-\rightarrow H^+H^-$ up to NNLO and the corresponding relative corrections as functions of $\tan\beta$.}
    \label{fig:tb}
\end{figure*}
In Fig.\ref{fig:tb}, we compare the different behaviors between results in type-I and type-II THDMs with respect to the change of $\tan\beta$ at different charged Higgs masses. For the $O(\alpha)$ relative corrections, the type-I results increase no more than $3.16\%$, while the type-II results decreases from $-2.07\%$ to $-16.00\%$ for $m_{H^{\pm}}=800\ \mathrm{GeV}$ and from $-18.02\%$ to $-27.86\%$ for $m_{H^{\pm}}=1200\ \mathrm{GeV}$. For the $O(\alpha\alpha_s)$ relative corrections, the type-I results decrease from $1.33\%$ to $0.96\%$ for $m_{H^{\pm}}=130\ \mathrm{GeV}$ and increase from $1.03\%$ to $1.12\%$ for $m_{H^{\pm}}=400\ \mathrm{GeV}$, while the type-II results increase from $1.17\%$ to $10.96\%$ for $m_{H^{\pm}}=800\ \mathrm{GeV}$ and from $1.07\%$ to $8.01\%$ for $m_{H^{\pm}}=1200\ \mathrm{GeV}$. The NNLO cross section differs small in the type-I THDM out of the suppressed vertex diagram contributions by $\tan\beta$, while for type-II at the two $m_{H^{\pm}}$ values because the variations of $O(\alpha)$ and $O(\alpha\alpha_s)$ relative corrections counteract with each other. These behaviors comply with the analyses of Yukawa couplings in type-I and type-II.

\par
\begin{figure*}
    \centering
    \includegraphics[scale=0.78]{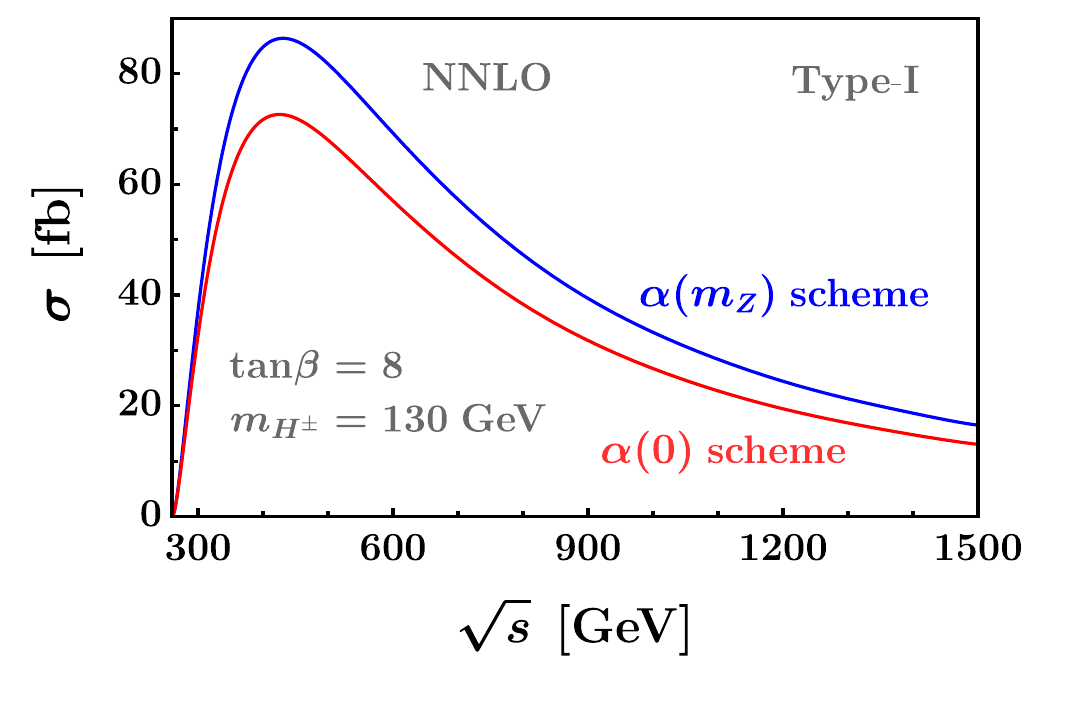}
    \includegraphics[scale=0.78,trim= 40pt 0 0 0]{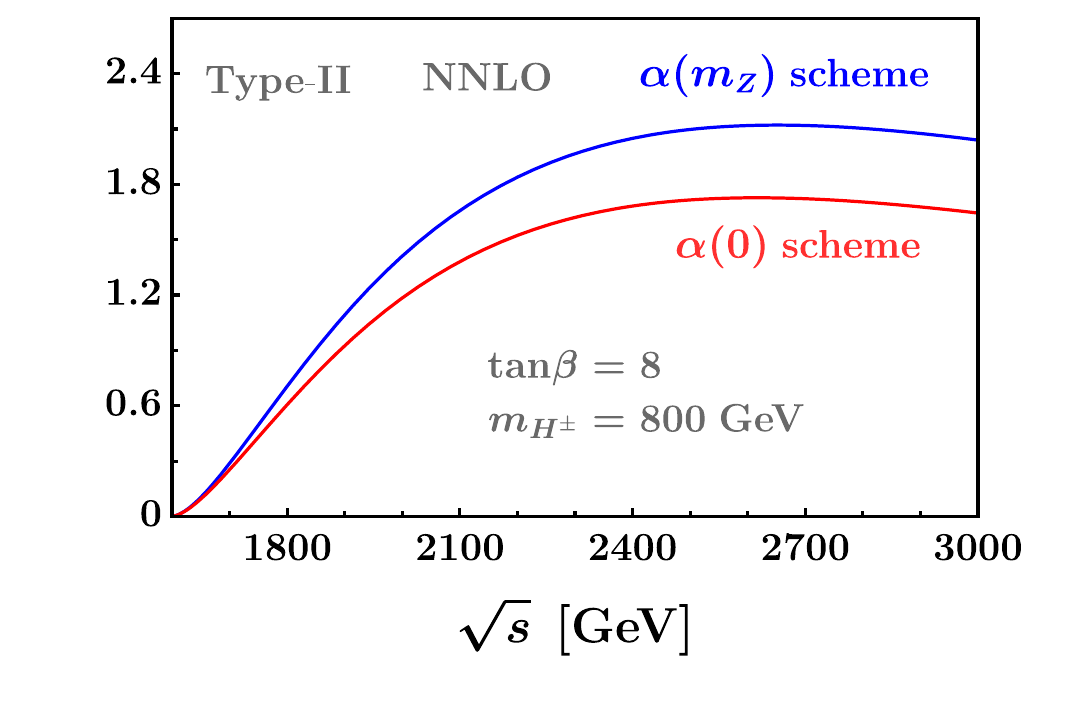}
	\caption{Corrected cross sections of $e^+e^-\rightarrow H^+H^-$ up to NNLO in different input parameter schemes.}
    \label{fig:alpha}
\end{figure*}
\begin{table*}[tbp]
    \centering
    \renewcommand{\arraystretch}{1.1}
      \begin{tabular}{ccccccccc}
        \toprule[1.2pt]
	      {Type} & {scheme} & $\sigma_{0}\ [\mathrm{fb}]$ & $ \delta_{\mathrm{EW}}\ [\%]$ & $\sigma_{\mathrm{NLO}}\ [\mathrm{fb}]$ & $\delta_{\mathrm{QCD}\otimes\mathrm{EW}}\ [\%]$ &  $\sigma_{\mathrm{NNLO}}\ [\mathrm{fb}]$ \\
              \hline
              \hline
	      \multirow{2}{*}{I}  & {$\alpha(m_Z)$} & {$13.01$} & $25.72$ &  $16.35$   & $1.11$ & $16.50$ \\
				  & {$\alpha(0)$}   & {$11.52$}   & $12.03$ &  $12.90$   & $1.04$ & $13.02$ \\
              \hline
              \multirow{2}{*}{II} & {$\alpha(m_Z)$} & {$2.06$}   & $-2.11$ &  $2.02$   & $1.22$ & $2.04$ \\
		                  & {$\alpha(0)$}   & {$1.82$}  & $-10.91$ &  $1.62$   & $1.15$ & $1.64$ \\
        \bottomrule[1.2pt]
      \end{tabular}    
	\caption{The results for type-I and type-II in $\alpha(0)$ and $\alpha(m_Z)$ schemes, where $\sigma_0$, $\sigma_{\mathrm{NLO}}$, and $\sigma_{\mathrm{NNLO}}$ are cross sections evaluated up to Born-level, NLO and NNLO respectively. $\delta$ represents the relative correction at the corresponding order. For type-I, $\sqrt{s}=1500\ \mathrm{GeV}$, $\tan\beta=8$ and $m_{H^{\pm}}=130\ \mathrm{GeV}$. For type-II, $\sqrt{s}=3000\ \mathrm{GeV}$, $\tan\beta=8$ and $m_{H^{\pm}}=800\ \mathrm{GeV}$.}
\label{tab:alpha}
\end{table*}
The NNLO integrated cross sections in $\alpha(0)$ and $\alpha(M_Z)$ schemes are depicted in Fig.\ref{fig:alpha}, whose behaviors are consistent. For type-I, they both reach the maximum near $\sqrt{s}=430 \ \mathrm{GeV}$, while the cross section in the $\alpha(M_Z)$ scheme is always larger and has a maximum of $86.4$ fb and that in the $\alpha(0)$ scheme is $72.6$ fb. For type-II, the cross section has a maximum of $2.12$ fb in the $\alpha(M_Z)$ scheme and $1.73$ fb in the $\alpha(0)$ scheme. In Tab.\ref{tab:alpha}, we list the results of the two schemes. The $O(\alpha\alpha_s)$ corrections of the two schemes only differ by a factor of $\alpha(0)/\alpha(m_Z)$ because of no correction to the $\Delta\alpha$ in this order. 

\par
\begin{figure*}
    \centering
    \includegraphics[scale=0.78]{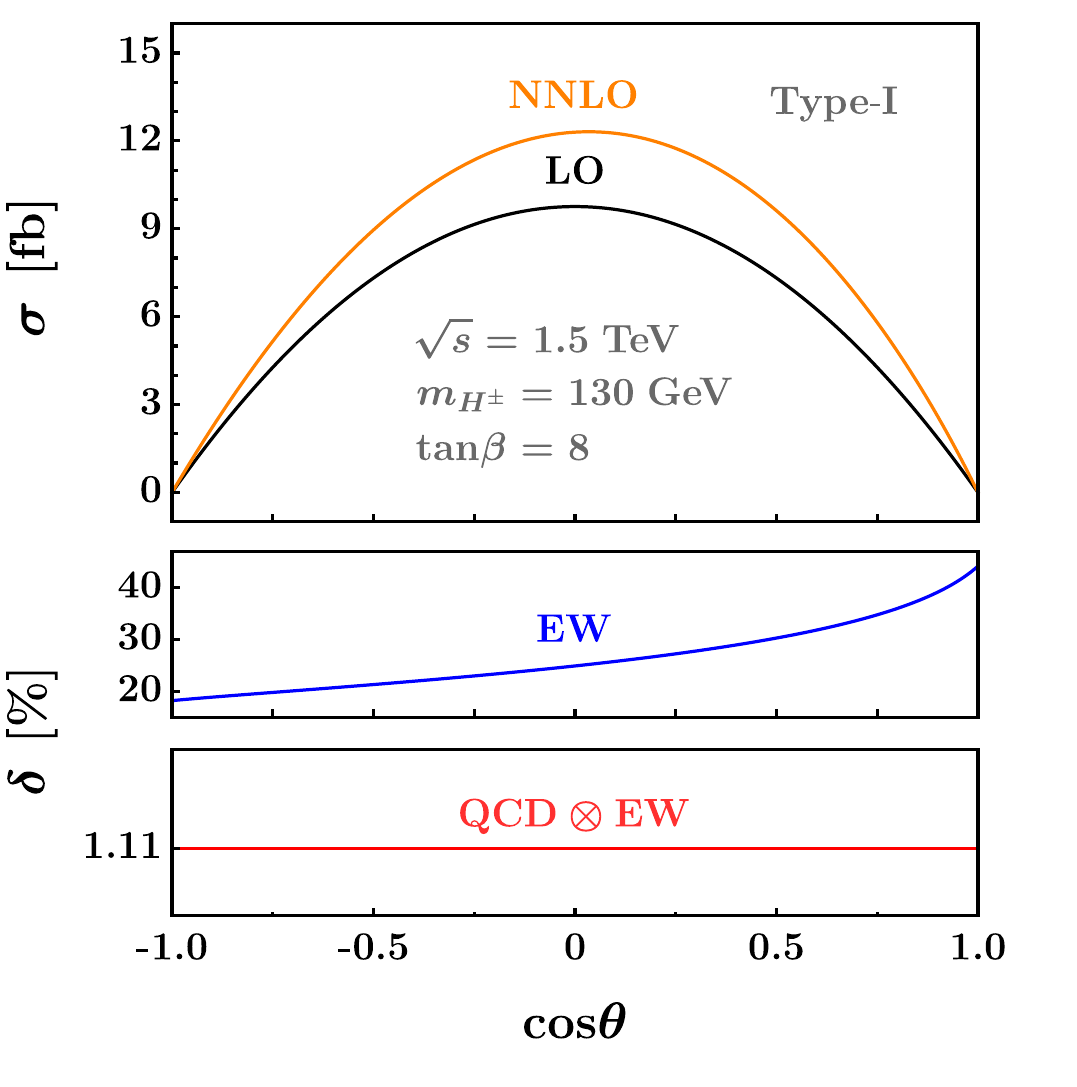}
    \includegraphics[scale=0.78,trim= 40pt 0 0 0]{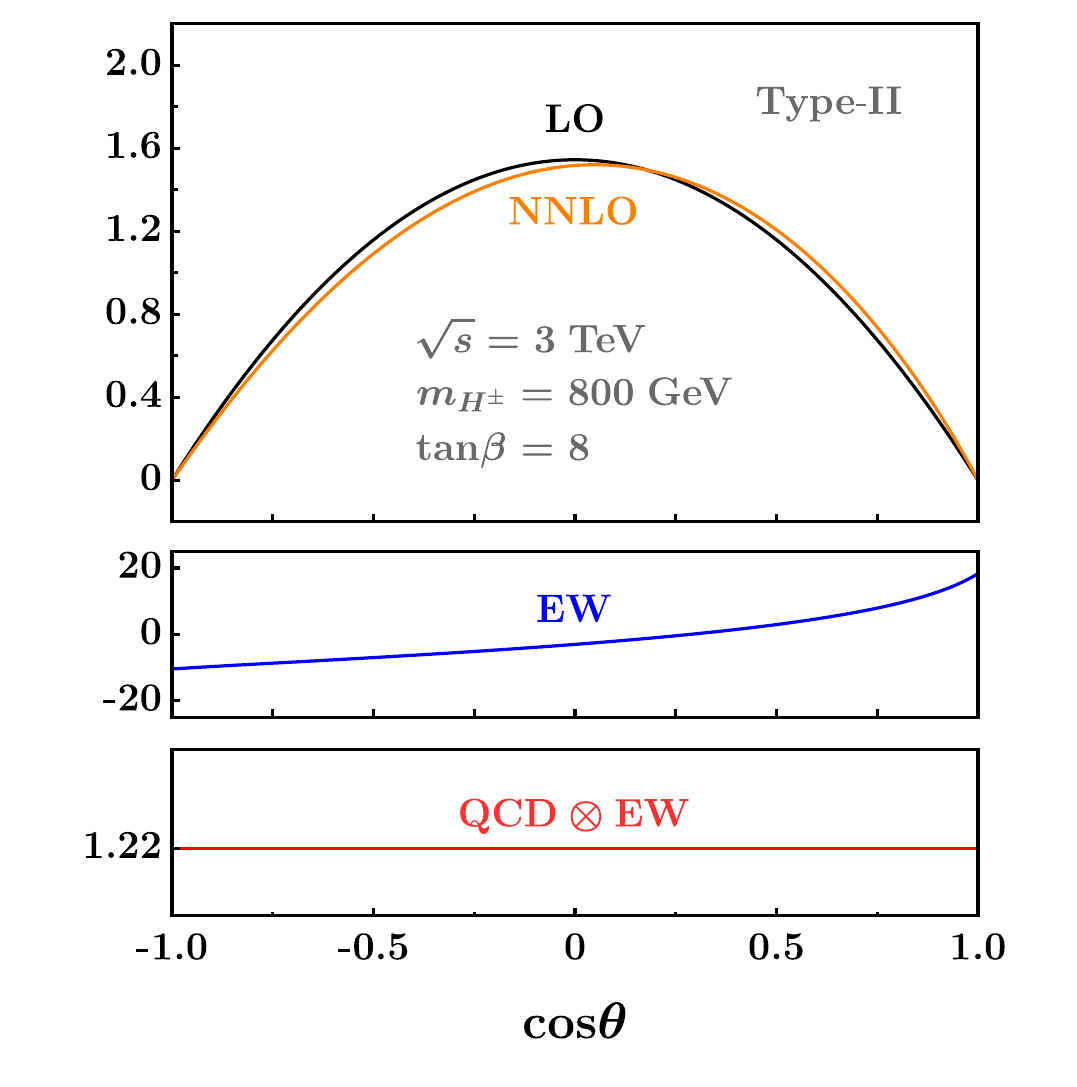}
	\caption{Corrected differential cross sections of $e^+e^-\rightarrow H^+H^-$ up to NNLO in association with Born differential cross sections and corresponding relative corrections as functions of the scattering angle.}
    \label{fig:dCS}
\end{figure*}
We plot differential cross sections up to LO and NNLO in Fig.\ref{fig:dCS}, along with the relative corrections of $O(\alpha)$ and $O(\alpha\alpha_s)$ as functions of the scattering angle. The curve representing the LO result is a quadratic function, as the LO cross section is proportional to $\sin^2\theta$. Notably, the $O(\alpha\alpha_s)$ correction is also proportional to $\sin^2\theta$, thus the $O(\alpha\alpha_s)$ relative correction has no dependence on $\cos\theta$, and we will explain it in \ref{sec:independence}.

\par
\begin{figure*}
    \centering
       \includegraphics[scale=0.52]{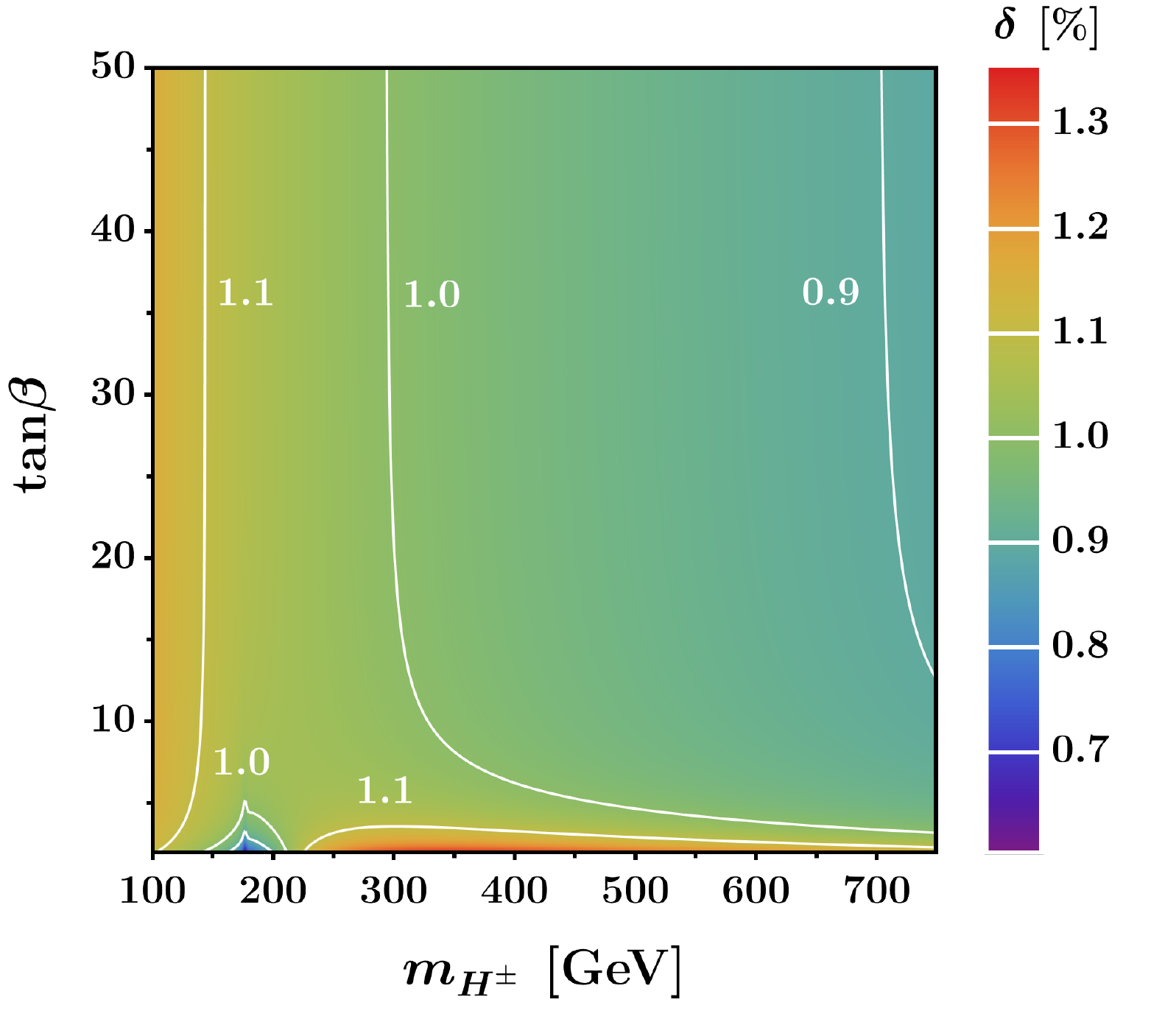}
       \includegraphics[scale=0.52,trim= 40pt 0pt 0pt 0pt]{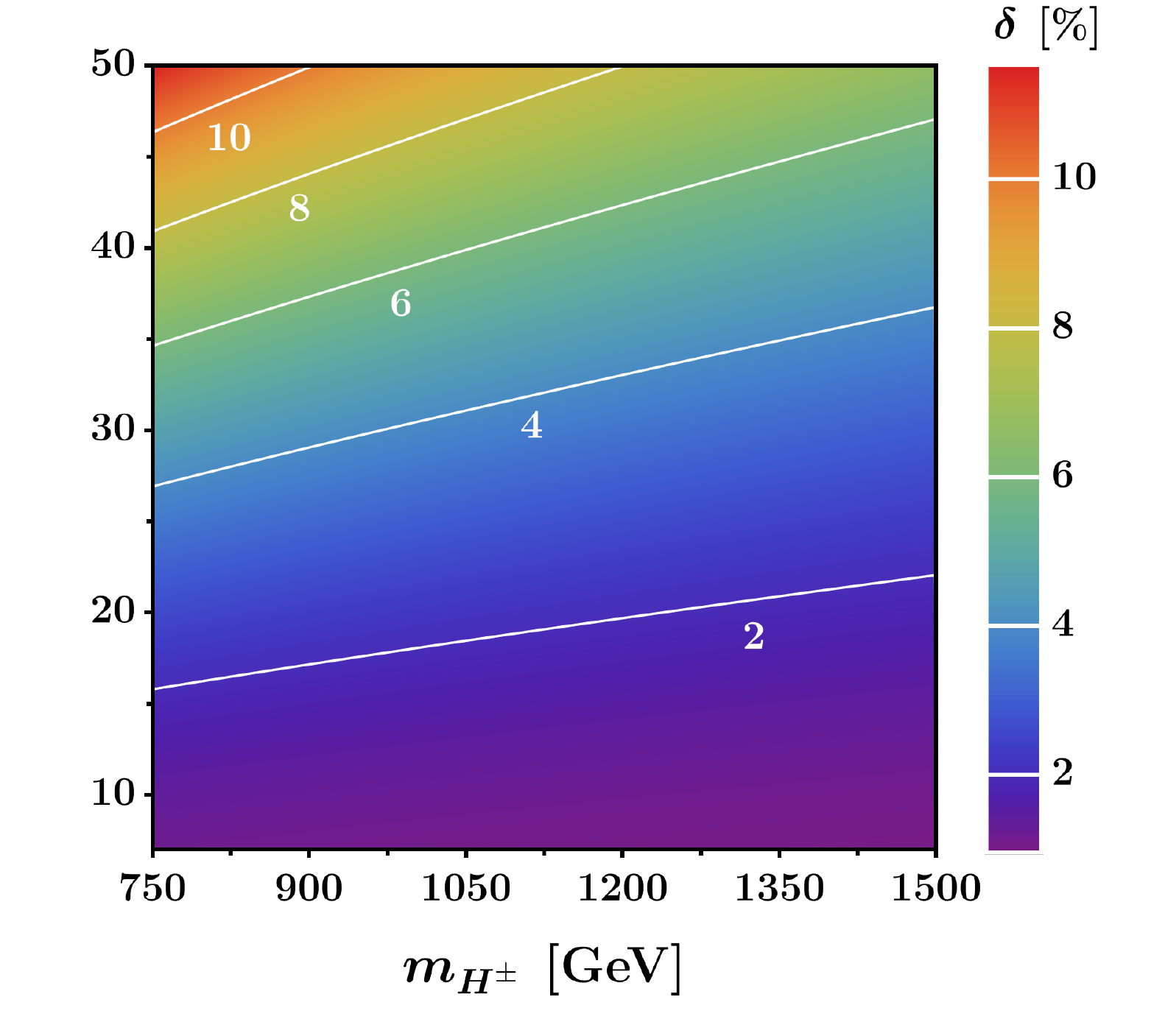}
	\caption{Relative corrections of $e^+e^-\rightarrow H^+H^-$ at NNLO in type-I (left) and type-II (right) THDMs.}
    \label{fig:contour}
\end{figure*}
In Fig.\ref{fig:contour}, we give the $O(\alpha\alpha_s)$ relative corrections in density plots. $m_{H^\pm}$ and $\tan\beta$ vary in the union of allowed regions of type-I (type-II) and type-X (type-Y): $\tan\beta\in [2, 50]$, $m_{H^\pm}\in [100,750]\ \mathrm{GeV}$ at $\sqrt{s}=1500 \ \mathrm{GeV}$ for type-I and type-X and $\tan\beta\in [7, 50]$, $m_{H^\pm}\in [750,1500]\ \mathrm{GeV}$ at $\sqrt{s}=3000 \ \mathrm{GeV}$ for type-II and type-Y, respectively. From the right panel, we can see that the $O(\alpha\alpha_s)$ relative correction in type-II can reach up to $11.5\%$ when $\tan\beta=50$, while the $O(\alpha\alpha_s)$ relative correction in type-I is suppressed by $\tan\beta$ and varies mainly in $\left[0.9\%,1.1\%\right]$, as the left panel shows.

\section{Summary}
\label{sec:summary}
\par
Searching for the charged Higgs boson is a task of significance for both present and future experiments, as it would be a conclusive signal of BSM physics. It is predicted by the THDM where abundant phenomenology is offered. With higher luminosity and cleaner background, future $e^+e^-$ colliders are in the ascendent to search for $H^{\pm}$, and the pair production process is the main channel for this purpose. In this work, we provide a detailed study on the $e^+e^- \rightarrow H^+H^-$ process within the framework of four tyes of THDMs. We review the experimental and analytical constraints on the THDM parameters, especially for $m_{H^{\pm}}$ and $\tan\beta$ that the $O(\alpha\alpha_s)$ correction depends on. Generally, the constraints for type-II and type-Y are much stricter than those for type-I and type-X. We investigate the $O(\alpha\alpha_s)$ corrections at different $m_{H^\pm}$, $\tan\beta$ and colliding energies, and provide numerical results for the NNLO differential and integrated cross sections. At $m_{H^\pm} = 130\ \mathrm{GeV}$ and $\tan\beta=8$, the $O(\alpha\alpha_s)$ relative correction in the type-I THDM increases slowly around $1.0\%$, with a resonance peak at $\sqrt{s}=2m_t$. The $O(\alpha\alpha_s)$ relative correction of type-II increases slowly around $1.1\%$ at $m_{H^\pm} = 800\ \mathrm{GeV}$ and $\tan\beta=8$. Another resonance peak appears at $m_{H^{\pm}}=m_t+m_b$ for both $O(\alpha)$ and $O(\alpha\alpha_s)$ corrections, and is suppressed by large $\tan\beta$ in the type-I THDM. From the Yukawa couplings of $H^{\pm}$ and quarks, one can conclude that the dependences of $O(\alpha\alpha_s)$ relative corrections on $\tan\beta$ are simple functions. In type-I THDM, the dependence is monotone, while in type-II THDM is not, and the amplification of large $\tan\beta$ on $O(\alpha\alpha_s)$ corrections for type-II is remarkable. In the given parameter region, the $O(\alpha\alpha_s)$ relative correction in type-I THDM varies mostly less than $0.1\%$ of absolute values from $1.0\%$, being relatively steady compared to the extensive range variation, from $1.01\%$ to $11.5\%$, in the type-II THDM. Given this correction magnitude, the $O(\alpha\alpha_s)$ relative correction is nonnegligible.

\par
Through calculations and analyses, we find that $O(\alpha\alpha_s)$ relative corrections of this process have no dependence on the scattering angle determined by the topology of Feynman diagrams. Despite numerous unfixed parameters in the THDM, the $O(\alpha\alpha_s)$ contribution is relevant with only two of them. Moreover, the $O(\alpha\alpha_s)$ correction in the $\alpha(0)$ scheme can be converted through a constant factor from that in the $\alpha(m_Z)$ scheme. Therefore, when studying this process with other varying parameters, such as $m_{A^0},m_{H^0},\sin\alpha$ and $\lambda_5$, the $O(\alpha\alpha_s)$ contribution acts as an invariant correction, especially with its independence on the scattering angle, thus being valid in various scenarios.

\par
We also present the $O(\alpha)$ correction and the NNLO corrected cross section for completeness. The corrected cross section is generally at tens of fb and can be larger than $80\ \mathrm{fb}$ in the allowed regions of type-I and type-X THDMs, having a sizeable potential to be detected in future $e^+e^-$ colliders. For type-II and type-Y, the corrected cross section is at several fb in the allowed region.

\vskip 5mm

\noindent{\large\bf Acknowledgments:}
\par
This work is supported by the National Natural Science Foundation of China (Grant No. 12061141005) and the CAS Center for Excellence in Particle Physics (CCEPP).

\appendix
\section{Independence on scattering angle}
\label{sec:independence}
\par
Consider an $e^+e^-$ initial state Feynman diagram where momenta $p_{1,2}$ appear only in spinor field operators $\bar{v}(p_1)$ and $u(p_2)$, which is true for a Feynman diagram without initial state vertex correction loop, in other words, where the initial state electron and positron couple with the exact single propagator. The above condition is established in the tree and two-loop mixed QCD$\otimes$EW diagrams in this process. As a result, the dependences of the Born cross section and $O(\alpha\alpha_s)$ corrections on $p_{1,2}$ come only from the Dirac traces generated by the interference between tree-level and two-loop amplitudes. As initial state vertices only include $e^+e^-\gamma$ and $e^+e^-Z$, nonzero Dirac traces can be simplified to the following form
\begin{equation}
	\sum_{spins}\bar{u}(p_2)\cancel{q}_2v(p_1)\bar{v}(p_1)\cancel{q}_1u(p_2) = tr(\cancel{p}_2\cancel{q}_2\cancel{p}_1\cancel{q}_1) 
\end{equation}
where $q_{1,2}$ are linear combinations of $p_3$ and $p_4$, $\sigma$ denotes the spin, and the notation $\cancel{p}\equiv\gamma^\mu p_\mu$. Furthermore, $q_{1,2}$ can also be written as linear combinations of $p_3+p_4$ and $p_3-p_4$, where the former part is proportional to the tiny $m_e$, by using the Dirac equation, and thus negligible. Consequently, the above nonzero traces are all proportional to 
\begin{equation}
	tr(\cancel{p}_2\cancel{p}_{34}\cancel{p}_1\cancel{p}_{34}) = 2s(s-4m_{H^\pm}^2)\sin^2\theta
\end{equation}
where $p_{34} \equiv p_3-p_4$. Remember that $\theta$ enters in the cross section via the product between initial and final states momenta $p_i\cdot p_f \ (i=1,2;\ f=3,4)$, so the Born cross section and the $O(\alpha\alpha_s)$ correction are proportional to $\sin^2\theta$ and have no other $\theta$ dependence.

\par
This conclusion could also be deduced for the diagrams of the initial state vertex correction topology in the $O(\alpha)$ with a similar analysis but could not be applied to box diagrams. Thus, the phenomena of forward-backward asymmetry appear in the $O(\alpha)$ correction.

\bibliographystyle{apsrev4-2}
\bibliography{ref}

\end{document}